\documentclass[10pt,article,oneside,oldfontcommands]{memoir}
\setsecnumdepth{subsubsection}

\usepackage{blindtext}
\usepackage{amsmath}
\usepackage{amsthm}
\usepackage{amssymb}
\usepackage{amsfonts}
\usepackage{graphicx}
\usepackage{mathrsfs}
\usepackage{tikz}
\usepackage{hyperref}
\usepackage[nospace]{cite}
\usepackage{authblk}
\usepackage{mathtools}
\usepackage{float}
\usepackage[Symbol]{upgreek}
\usepackage{relsize}
\usepackage{wrapfig}
\usepackage{setspace}

\settypeblocksize{0.75\stockheight}{0.7\stockwidth}{*}
\setlrmargins{*}{*}{1}
\setulmargins{*}{*}{1.5}
\checkandfixthelayout[nearest]

\theoremstyle{mytheoremstyle}

\theoremstyle{mytheoremstyle}

\theoremstyle{mytheoremstyle}

\theoremstyle{theorem}

\theoremstyle{definition}

\theoremstyle{remark}

\theoremstyle{theorem}

\def\bs{\boldsymbol}
\def\d{\mathrm{d}}
\def\i{\mathrm{i}}
\def\e{\mathrm{e}}

\counterwithout{section}{chapter}

\title{Anyon Chains with Pairing Terms}
\author[1,2,3]{B. Majidzadeh Garjani}
\author[1]{E. Ardonne}
\affil[1]{Department of Physics, Stockholm University, SE-106 91 Stockholm, Sweden}
\affil[2]{Department of Mathematics, Stockholm University, SE-106 91 Stockholm, Sweden}
\affil[3]{Nordita, Royal Institute of Technology and Stockholm University, Roslagstullsbacken 23, SE-10691 Stockholm, Sweden}

\newcommand\raisepunct[1]{\mathpunct{\raisebox{0.45ex}{#1}}}

\begin{document}

\maketitle

\begin{abstract}
In this paper we introduce a one-dimensional model of $su(2)_k$ anyons in which the number
of anyons can fluctuate by means of a pairing term.
The model can be tuned to a point at which one can determine the exact zero-energy ground states,
in close analogy to the spin-1 AKLT model. We also determine the points at which the model is
integrable and determine the behavior of the model at these integrable points.
\end{abstract}


\section{Introduction}

Almost forty years ago, it was realized by Leinaas and Myrheim that in two-dimensional
systems, the existence of particles with statistics interpolating between bosonic and fermionic statistics is a possibility
\cite{leinaas}. It is widely believed that particles of this type are realized in the fractional
quantum Hall systems \cite{tsui,laughlin83}, even though the anyonic statistics of the quasi-particles
has not yet been probed directly. Moore and Read proposed fractional quantum Hall states  for which
the quasi-particles have non-Abelian statistics \cite{MooreRead}. It is believed that this Moore--Read
state describes the $\nu=5/2$ quantum Hall effect \cite{morf}. Interestingly, the one-dimensional
$p$-wave superconductor studied by Kitaev \cite{kitaev-chain} exhibits Majorana bound states,
which are the one-dimensional cousins of the non-Abelian anyons present in the Moore--Read state.
Following the first theoretical proposal \cite{oreg,lutchyn} of how to realize Kitaev's model, there
are experimental indications that Majorana bound states might be realized experimentally
\cite{mourik,deng,das}.

In \cite{feiguin}, the effect of interactions between non-Abelian anyons were studied by means
of a one-dimensional model-Hamiltonian.
Non-Abelian anyons, in particular so-called Fibonacci anyons,
are used as building blocks for this model-Hamiltonian 
in the same way as spins are used in model-Hamiltonians such as the Heisenberg model in order
to study magnetism. In \cite{fibintro} different types of anyon models are explained, while
\cite{lesanovsky12} proposes an experimental realization of interacting Fibonacci anyons.

Various generalizations of the original model have already been considered. These include
models with longer-range interactions \cite{anyon-long-range}, models on ladders \cite{gils,schultz,soni}, as well as models making use of different types of anyons\cite{spin1,gils2,dancer,finch}.
In this paper, we consider a generalization of the
dilute anyon model \cite{dilute}---a model in which anyons are allowed to hop
to empty neighboring sites. The type of term we add to this model is a pairing term that creates or annihilates pairs of anyons on neighboring sites. The anyons
we use are of the type $su(2)_k$, which is the same type as the excitations of the
Read--Rezayi quantum Hall states \cite{rr99}.

The anyon model we study has a large number of parameters and, since the number of anyons is not considered to be fixed in our model, the size of the Hilbert space
grows quickly with system size. We
therefore can not characterize the phase diagram of our model in full detail, but we concentrate
ourselves on two particular cases. In the first case, we tune the model to a point where the
Hamiltonian becomes a sum of projectors. This allows us to determine the exact zero-energy
ground states for the case when $k$, in $su(2)_k$,  is odd. In addition, we study the model at
two integrable points where, in most cases,  the model turns out to be critical. Most of these
critical points are described by minimal-model conformal field theories.

The outline of this paper is as follows.
In Section \ref{sec:anyon-theory}, we introduce the notion of anyon models as the language
we use to define our Hamiltonian.  In Section \ref{sec:model}, we introduce the anyon model and its corresponding Hilbert space and finally present the Hamiltonian that we investigate in later sections. 
In Section \ref{sec:exact-groundstates}, we study the Hamiltonian at a special point where we can
determine the exact zero-energy ground states as well as the number of these states.
In Section \ref{sec:integrability}, we determine for which values of the parameters the model is integrable 
and determine the behavior of the model at these integrable points.
Section \ref{sec:conclusions} is devoted to the conclusions.
In Appendix \ref{app:f-symbols}, we give the explicit form of the $F$-symbols we use in this paper.
 In Appendix \ref{app:k1alternative}, we map our model-Hamiltonian, in the case of $k=1$, to a spin-1/2
Hamiltonian.


\section{General Theory of Anyons}
\label{sec:anyon-theory}
The mathematical framework that describes the anyons in a rigorous way is that of tensor categories.
However, in this article we do not need this full machinery and we use a more concrete picture of
anyons and anyon systems. We will be brief here and refer to
\cite{Kitaev06,bonderson-thesis,Wang} for more elaborate introductions into the subject.

We start introducing a finite set of \emph{labels}, $\mathcal{L} = \{a,b,c,\ldots,n\}$, that contains the labels of all the
anyon types present in the anyon model we consider. We call these labels the anyon `\emph{charges}'.
One of the elements of this set is distinguished
from others and plays the role of the \emph{vacuum}. We label this element by $\bs{1}$.
In addition, with each label $a$ in the set, we associate another label in $\mathcal{L}$, denoted by
$\widehat{a}$, that represents the dual of $a$.
The dual to the vacuum is the vacuum itself, that is, $\widehat{\bs{1}} = \bs{1}$. For other anyons $a$, the dual might
or might not be the original anyon, but we always have $\widehat{\widehat{a}} = a$. To define the
notions of the vacuum and the dual anyon, we need to introduce the notion of \emph{fusion} first.

The fusion of anyons is analogous to combining different spin multiplets by means of the
tensor product. Thus, to specify the possible fusions, we need to specify what the possible `fusion outcomes' are for each pair of
anyons $a$ and $b$. Symbolically, we write this as
\begin{equation}
\label{eq:fusionrule}
a \otimes b = \bigoplus_{c\in \mathcal{L}} N_{ab}^c\, c,
\end{equation}
where the fusion \emph{coefficients} $N_{ab}^c$ are non-negative integers. If $N_{ab}^c \geqslant 1$, this means that the overall charge of an anyon of type $a$ and an
anyon of type $b$ can be an anyon of type $c$, while this  possibility is ruled out if $N_{ab}^c = 0$. The labels $c$ for which $N_{ab}^c\geqslant 1$ are called fusion \emph{channels} of $a$ and $b$ and if the outcome of the fusion of $a$ and $b$ is $c$, then $c$ is said to be the fusion channel of $a$ and $b$. 
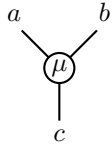
\begin{figure}
\centering
\begin{tikzpicture}
    \draw [thick] (0,0) circle (.2cm);
	\draw [thick] (0,-.7) -- (0,-0.2);
	\draw [thick] (-.1414,.1414) -- (-.5,.5);
	\draw [thick] (.1414,.1414) -- (.5,.5);
	\node [left,above] at (-.6,.5) {$a$};
	\node [right,above] at (.6,.5) {$b$};
	\node [below] at (0,-.7) {$c$};
	\node at (0,0) {$\mu$};
\end{tikzpicture}\caption{Anyons $a$ and $b$ fuse to anyon $c$ in the $\mu$th way.}\label{fig:fuse}
\end{figure}
\noindent If
$N_{ab}^c > 1$, then this means that anyons $a$ and $b$ can be fused to the anyon-type $c$ in more than one way. The fusion of charges $a$ and $b$ that has given rise to charge $c$ in the $\mu$th way,
$\mu=1,2,\ldots,N_{ab}^c$, is usually depicted graphically as in Figure \ref{fig:fuse} and it is called a fusion \emph{tree}. 

 The anyon models we consider in this paper are \emph{multiplicity-free} models, that is, they have the property that $N_{ab}^c$ is either zero or one for all labels $a$, $b$, and $c$ in $\mathcal{L}$. We therefore
can simply omit the label $\mu$ in the remainder of the paper.

We now specify the physical constraints on the fusion rules:
\begin{enumerate}[(i)]
\item
The vacuum $\bs{1}$ is the
unique label such that $N_{\bs{1}a}^c = \delta_{ac}$, for all labels $a$ and $c$ in $\mathcal{L}$, with $\delta$ denoting the Kronecker delta.
\item
The dual of $a$, namely $\widehat{a}$, 
is the unique label such that $N_{a b}^{\bs{1}} = \delta_{b\widehat{a}}$, which means
that $a$ and $\widehat{a}$ can be fused to the vacuum.
\item
The fusion rules are \emph{associative} in the sense that the set of all possible fusion outcomes of
$a\times (b\times c)$ is equal to the set of all possible fusion outcomes of
$(a\times b)\times c$. In terms of the fusion coefficients, this means that
\begin{equation}
\sum_{e\in \mathcal{L}}N_{ab}^e\,N_{ec}^d=\sum_{f\in \mathcal{L}}N_{bc}^f\,N_{af}^d,
\end{equation}
for any labels $a$, $b$, $c$, and $d$ in $\mathcal{L}$.
\item
Finally, we demand that fusion is `symmetric' in the following sense:
\begin{equation} 
N_{ab}^c = N_{ba}^c = N_{b\widehat{c}}^{\widehat{a}} = N_{\widehat{a}\widehat{b}}^{\widehat{c}}\,.
\end{equation}
\end{enumerate}

An anyon theory is said to be \emph{non-Abelian} if there are labels $a$ and $b$ such that 
$\sum_{c\in \mathcal{L}}N_{ab}^c>1$, otherwise, it is called \emph{Abelian}. The Fibonacci anyon theory, described in the following, is an example of a non-Abelian and multiplicity-free anyon theory. The label set of this model is $\mathcal{L}=\{\bs{1},\tau\}$, where $\tau$, known as the \emph{Fibonacci} anyon, is the only non-trivial anyon of the model and the fusion rules are given by 
\begin{align*}
   \bs{1}\otimes\bs{1}&=\bs{1},\\
   \tau\otimes\bs{1}&= \bs{1} \otimes \tau = \tau,\\
   \tau\otimes\tau &=\bs{1}\oplus\tau.
\end{align*}
From the last fusion rule it follows that the Fibonacci anyon is its own dual.

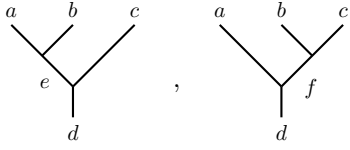
\begin{figure}[h]
\centering\begin{tikzpicture}[baseline, scale=0.82, every node/.style={scale=0.82}]
		\draw [thick] (0,0)--(-1,1);
		\draw [thick] (0,0)--(1,1);
		\draw [thick] (0,0)--(0,-.5);
		\draw [thick] (-0.5,0.5)--(0,1);
		\node [below] at (0,-.5) {$d$};
		\node [above] at (-1,1) {$a$};
		\node [above] at (0,1) {$b$};
		\node [above] at (1,1) {$c$};
		\node [below left] at (-.25,.25) {$e$};
	\end{tikzpicture}\quad,\quad
	\begin{tikzpicture}[baseline, scale=0.82, every node/.style={scale=0.82}]
		\draw [thick] (0,0)--(-1,1);
		\draw [thick] (0,0)--(1,1);
		\draw [thick] (0,0)--(0,-.5);
		\draw [thick] (0.5,0.5)--(0,1);
		\node [below] at (0,-.5) {$d$};
		\node [above] at (-1,1) {$a$};
		\node [above] at (0,1) {$b$};
		\node [above] at (1,1) {$c$};
		\node [below right] at (.25,.25) {$f$};
	\end{tikzpicture}\caption{Possible fusion trees for anyons $a$, $b$, and $c$ fusing to $d$.}\label{fig:basic fusion trees}
\end{figure}

To be able to construct a model describing anyons that can interact with each other, we have to
associate a Hilbert space with a collection of anyons. We do this in more detail in the following
section. Here, however, we concentrate ourselves on up to four anyons in order to explain the
so-called \emph{Pentagon} equations.
With two anyons $a$ and $b$ with fusion
channel $c$, we associate the Hilbert space $\mathcal{H}_{ab}^c$ of dimension
$N_{ab}^c$. We can label the basis-elements of this space by the diagrams depicted in Figure \ref{fig:fuse}, excluding the label $\mu$ for the reason mentioned earlier.
For the case of three anyons $a$, $b$, and $c$, with the overall fusion channel $d$, we denote
the associated Hilbert space by $\mathcal{H}_{abc}^{d}$. In this case, we can consider two different bases.
Either $a$ fuses with $b$ first and then the outcome $e$ fuses with $c$ to give $d$, or $b$ fuses
with $c$ first, to give $f$, which fuses with $a$ to give $d$.
These possibilities are shown in Figure \ref{fig:basic fusion trees}. The dimension of this
Hilbert space is given by
$\sum_{e\in \mathcal{L}}N_{ab}^e\,N_{ec}^d = \sum_{f\in \mathcal{L}}N_{bc}^f\,N_{af}^d$, where the equality
follows from the associativity of the fusion rules.

The two different ways of describing the Hilbert space $\mathcal{H}_{abc}^d$ are related by
a basis transformation. The basis-transformation matrix
describing this basis transformation is called the $F$-\emph{matrix} and it is denoted by $F^{abc}_d$. 
Going between the fusion tree on the left  to the fusion tree on right in Figure \ref{fig:basic fusion trees},
is called an $F$-\emph{move}. When dealing with physical theories,  the $F$-matrices are demanded to be invertible and unitary. Symbolically we write
\begin{align}\label{eq:fequation}
	\begin{tikzpicture}[baseline, scale=0.82, every node/.style={scale=0.82}]
		\draw [thick] (0,0)--(-1,1);
		\draw [thick] (0,0)--(1,1);
		\draw [thick] (0,0)--(0,-.5);
		\draw [thick] (-0.5,0.5)--(0,1);
		\node [below] at (0,-.5) {$d$};
		\node [above] at (-1,1) {$a$};
		\node [above] at (0,1) {$b$};
		\node [above] at (1,1) {$c$};
		\node [below left] at (-.25,.25) {$e$};
	\end{tikzpicture}\hspace{-.5cm}=\sum_{f} F^{abc}_{d;\,ef}\hspace{-.5cm}
	\begin{tikzpicture}[baseline, scale=0.82, every node/.style={scale=0.82}]
		\draw [thick] (0,0)--(-1,1);
		\draw [thick] (0,0)--(1,1);
		\draw [thick] (0,0)--(0,-.5);
		\draw [thick] (0.5,0.5)--(0,1);
		\node [below] at (0,-.5) {$d$};
		\node [above] at (-1,1) {$a$};
		\node [above] at (0,1) {$b$};
		\node [above] at (1,1) {$c$};
		\node [below right] at (.25,.25) {$f$};
	\end{tikzpicture},
\end{align}
\noindent where the sum runs through all possible labels $f$ in the fusion channel of $b$ and $c$
such that $d$ is in the fusion channel of $a$ and $f$. The symbol 
$F^{abc}_{d;ef}$ denotes the $(e,f)$th entry of the matrix $F^{abc}_d$ and it is
called an $F$-\emph{symbol}.

As it stands, Equation~\eqref{eq:fequation} just means that an $F$-matrix
is a  basis-transformation matrix and does not constrain the $F$-symbols any further.
To determine the $F$-symbols, one needs to consider the Hilbert space associated
with four anyons. Two different bases for the Hilbert space $\mathcal{H}_{abc}^d$ are depicted by the
following two fusion trees:
\begin{center}
\begin{tikzpicture}[baseline, scale=0.82, every node/.style={scale=0.82}]
	\draw [thick] (0,0) -- (0,-.5);
	\draw [thick] (0,0) -- (-1.5,1.5);
	\draw [thick] (0,0) -- (1.5,1.5);
	\draw [thick] (-1,1) -- (-.5,1.5);
	\draw [thick] (-.5,.5) -- (.5,1.5);
	\node [above] at (-1.5,1.5) {$a$};
	\node [above] at (-.5,1.5) {$b$};
	\node [above] at (.5,1.5) {$c$};
	\node [above] at (1.5,1.5) {$d$};
	\node [below] at (0,-.5) {$e$};
	\node [below left] at (-.25,.25) {$v$};
	\node [below left] at (-.75,.75) {$u$};
\end{tikzpicture}\quad,\quad
\begin{tikzpicture}[baseline, scale=0.82, every node/.style={scale=0.82}]
	\draw [thick] (0,0) -- (0,-.5);
	\draw [thick] (0,0) -- (-1.5,1.5);
	\draw [thick] (0,0) -- (1.5,1.5);
	\draw [thick] (1,1) -- (.5,1.5);
	\draw [thick] (.5,.5) -- (-.5,1.5);
	\node [above] at (-1.5,1.5) {$a$};
	\node [above] at (-.5,1.5) {$b$};
	\node [above] at (.5,1.5) {$c$};
	\node [above] at (1.5,1.5) {$d$};
	\node [below] at (0,-.5) {$e$};
	\node [below right] at (.75,.75) {$y$};
	\node [below right] at (.25,.25) {$x$};
\end{tikzpicture}.
\end{center}
\noindent
One can describe the basis transformation between these two bases in two different ways, which
have to be equivalent to one another. The first one involves two $F$-moves:
\begin{align*}
\begin{tikzpicture}[baseline, scale=0.82, every node/.style={scale=0.82}]
	\draw [thick] (0,0) -- (0,-.5);
	\draw [thick] (0,0) -- (-1.5,1.5);
	\draw [thick] (0,0) -- (1.5,1.5);
	\draw [thick] (-1,1) -- (-.5,1.5);
	\draw [thick] (-.5,.5) -- (.5,1.5);
	\node [above] at (-1.5,1.5) {$a$};
	\node [above] at (-.5,1.5) {$b$};
	\node [above] at (.5,1.5) {$c$};
	\node [above] at (1.5,1.5) {$d$};
	\node [below] at (0,-.5) {$e$};
	\node [below left] at (-.25,.25) {$v$};
	\node [below left] at (-.75,.75) {$u$};
\end{tikzpicture}\xrightarrow{\text{$F$-move}}
\begin{tikzpicture}[baseline, scale=0.82, every node/.style={scale=0.82}]
	\draw [thick] (0,0) -- (0,-.5);
	\draw [thick] (0,0) -- (-1.5,1.5);
	\draw [thick] (0,0) -- (1.5,1.5);
	\draw [thick] (-1,1) -- (-.5,1.5);
	\draw [thick] (1,1) -- (.5,1.5);
	\node [above] at (-1.5,1.5) {$a$};
	\node [above] at (-.5,1.5) {$b$};
	\node [above] at (.5,1.5) {$c$};
	\node [above] at (1.5,1.5) {$d$};
	\node [below] at (0,-.5) {$e$};
	\node [below left] at (-.5,.5) {$u$};
	\node [below right] at (.5,.5) {$y$};
\end{tikzpicture}\xrightarrow{\text{$F$-move}}
\begin{tikzpicture}[baseline, scale=0.82, every node/.style={scale=0.82}]
	\draw [thick] (0,0) -- (0,-.5);
	\draw [thick] (0,0) -- (-1.5,1.5);
	\draw [thick] (0,0) -- (1.5,1.5);
	\draw [thick] (1,1) -- (.5,1.5);
	\draw [thick] (.5,.5) -- (-.5,1.5);
	\node [above] at (-1.5,1.5) {$a$};
	\node [above] at (-.5,1.5) {$b$};
	\node [above] at (.5,1.5) {$c$};
	\node [above] at (1.5,1.5) {$d$};
	\node [below] at (0,-.5) {$e$};
	\node [below right] at (.75,.75) {$y$};
	\node [below right] at (.25,.25) {$x$};
\end{tikzpicture},
\end{align*}
\noindent
and the second one involves three $F$-moves:
\begin{align*}
\begin{tikzpicture}[baseline, scale=0.82, every node/.style={scale=0.82}]
	\draw [thick] (0,0) -- (0,-.5);
	\draw [thick] (0,0) -- (-1.5,1.5);
	\draw [thick] (0,0) -- (1.5,1.5);
	\draw [thick] (-1,1) -- (-.5,1.5);
	\draw [thick] (-.5,.5) -- (.5,1.5);
	\node [above] at (-1.5,1.5) {$a$};
	\node [above] at (-.5,1.5) {$b$};
	\node [above] at (.5,1.5) {$c$};
	\node [above] at (1.5,1.5) {$d$};
	\node [below] at (0,-.5) {$e$};
	\node [below left] at (-.25,.25) {$v$};
	\node [below left] at (-.75,.75) {$u$};
\end{tikzpicture}\xrightarrow{\text{$F$-move}}
\begin{tikzpicture}[baseline, scale=0.82, every node/.style={scale=0.82}]
	\draw [thick] (0,0) -- (0,-.5);
	\draw [thick] (0,0) -- (-1.5,1.5);
	\draw [thick] (0,0) -- (1.5,1.5);
	\draw [thick] (-.5,.5) -- (.5,1.5);
	\draw [thick] (0,1) -- (-.5,1.5);
	\node [above] at (-1.5,1.5) {$a$};
	\node [above] at (-.5,1.5) {$b$};
	\node [above] at (.5,1.5) {$c$};
	\node [above] at (1.5,1.5) {$d$};
	\node [below] at (0,-.5) {$e$};
	\node [above left] at (-.25,.75) {$w$};
	\node [below left] at (-.25,.25) {$v$};
\end{tikzpicture}\xrightarrow{\text{$F$-move}}
\begin{tikzpicture}[baseline, scale=0.82, every node/.style={scale=0.82}]
	\draw [thick] (0,0) -- (0,-.5);
	\draw [thick] (0,0) -- (-1.5,1.5);
	\draw [thick] (0,0) -- (1.5,1.5);
	\draw [thick] (0,1) -- (.5,1.5);
	\draw [thick] (0,1) -- (-.5,1.5);
	\draw [thick] (.5,.5) -- (0,1);
	\node [above] at (-1.5,1.5) {$a$};
	\node [above] at (-.5,1.5) {$b$};
	\node [above] at (.5,1.5) {$c$};
	\node [above] at (1.5,1.5) {$d$};
	\node [below] at (0,-.5) {$e$};
	\node [below right] at (.25,.25) {$x$};
	\node [above right] at (.25,.75) {$w$};
\end{tikzpicture}\xrightarrow{\text{$F$-move}}
\begin{tikzpicture}[baseline, scale=0.82, every node/.style={scale=0.82}]
	\draw [thick] (0,0) -- (0,-.5);
	\draw [thick] (0,0) -- (-1.5,1.5);
	\draw [thick] (0,0) -- (1.5,1.5);
	\draw [thick] (1,1) -- (.5,1.5);
	\draw [thick] (.5,.5) -- (-.5,1.5);
	\node [above] at (-1.5,1.5) {$a$};
	\node [above] at (-.5,1.5) {$b$};
	\node [above] at (.5,1.5) {$c$};
	\node [above] at (1.5,1.5) {$d$};
	\node [below] at (0,-.5) {$e$};
	\node [below right] at (.75,.75) {$y$};
	\node [below right] at (.25,.25) {$x$};
\end{tikzpicture}.
\end{align*}
\noindent
This gives rise to the consistency conditions
\begin{align}\label{eqn:pentagon}
	F^{ucd}_{e;\,vy}\,F^{aby}_{e;\,ux}=\sum_{w}F^{abc}_{v;\,uw}\,F^{awd}_{e;\,vx}\,F^{bcd}_{x;\,wy},
\end{align} 
which are known as Pentagon equations. Here the sum is over all labels $w$ consistent with fusion rules. Although the Pentagon equations are obtained by considering only four anyons, Mac Lane's Coherence Theorem \cite{maclane} asserts that the Pentagon equations are all one needs to guarantee
consistency in the case of more than four anyons.
One should note that the Pentagon equations are just polynomial equations for $F$-symbols and it might be the case that there is no solution or there are several ones. It has been shown,
however, that the number of inequivalent solutions is actually finite. See Appendix \ref{app:f-symbols} for the notion
of gauge equivalence of $F$-symbols.
In this paper, a set of labels with consistent fusion rules together with a particular solution of
the Pentagon equations that leads to invertible $F$-matrices, is called an {\em anyon system}.

To completely describe an anyon system, one has to allow for the possibility of two anyons to be interchanged or \emph{braided}. The braiding of two anyons should also be consistent with fusion rules. This consistency gives rise to a set of equations called the {\em Hexagon} equations. In this paper, we do not consider the braiding of anyons  and refer the reader to \cite{Kitaev06,bonderson-thesis,Wang} for more details.


\section{Introducing the Anyon Model}
\label{sec:model}

In this section, we introduce the anyon model that we are interested in.
We briefly specify  the anyon system we are going to use, followed by a description of
the associated Hilbert space. Finally, we introduce the Hamiltonian that we study in this paper.  

\subsection{The Anyon System of the Model}
In this article, we are interested in the anyons with $su(2)_k$ fusion rules. These fusion rules
have close resemblance  to the ordinary $SU(2)$ tensor products. To express the $su(2)_k$ fusion
rules, one introduces a highest `spin' $k/2$, meaning that there are $k+1$ anyon types $0$, $1/2$, $1$, \ldots, $k/2$. Let $\mathcal{L}_k$ denote the set $\{0,1/2,\ldots,k/2\}$ of anyon types. Because of the presence of the `highest spin', the ordinary
$SU(2)$ tensor-product rules have to be modified that results in the following multiplicity-free associative
fusion rules:
\begin{align}\label{eq:fusion}
	i \otimes j = |i-j|\oplus(|i-j|+1)\oplus\cdots\oplus\min\{i+j,k-i-j\}\raisepunct{,}
\end{align}
for all $i$ and $j$ in $\mathcal{L}_k$. 
As mentioned in the previous section, the fusion rules do not, in general, fix the
$F$-symbols. Hence, we need to specify which set of the $F$-symbols we consider. For the $su(2)_k$ fusion rules, solutions to the Pentagon equations are known explicitly \cite{kr88,pasquier}. In Appendix~\ref{app:f-symbols}, we explicitly specify the form of the $F$-symbols we use to define the model .

Before we introduce the Hilbert space to our specific model, we first introduce the
\emph{quantum dimension} associated with an anyon type. The quantum dimension $d_a$ associated with a collection of anyons of type $a$, 
describes how the dimension of the Hilbert space corresponding to this collection of anyons  grows
with the number of such anyons. This dimension grows as $d_a^n$, where
$n$ is the number of anyons of type $a$. One can show \cite{Kitaev06} that the quantum dimensions
satisfy the following relation:
\begin{align}
	d_a^{\phantom{c}}d_b^{\phantom{c}}=\sum_{c \in\mathcal{L}_k} N_{a b}^{c}\, d_c^{\phantom{c}},
\end{align}
in resemblance with Equation~\eqref{eq:fusionrule}. In the case of $su(2)_k$ anyons, the quantum dimension
$d_j(k)$ of anyons of type $j$, for every $j$ in $\mathcal{L}_k$, is given by:
\begin{align}\label{eq:ds}
	d_j(k)=\frac{\sin\big(\frac{2j+1}{k+2}\,\uppi\big)}{\sin\big(\frac{1}{k+2}\,\uppi\big)}\raisepunct{.}
\end{align}
 Since $k$ remains fixed throughout the investigation of the model, we suppress $k$ dependence from the notation for these numbers and denote them simply by $d_j$. From Equation~\eqref{eq:ds}, one can immediately see that  $d_j=d_{k/2-j}$, for all labels $j$.

\subsection{The Hilbert Space of the Model}\label{subsec:hilbert space}

We want to model a physical system of anyons with $su(2)_k$ fusion rules in which the number of anyons can
fluctuate. We do this in the simplest possible setting, namely, we consider a chain consisting
of $l$ sites in which each site can be either occupied with an anyon of type $1/2$ or be empty. Besides, we do not
allow a site to be doubly occupied. A chain with all its sites occupied is called a \emph{dense} chain and one with some empty sites is called a \emph{dilute} chain.

First, one needs to introduce the Hilbert space of the model  and we do this by introducing an orthonormal basis for it. To do so, we need a set
of labels that keeps track of the occupation of the sites. We denote these labels by
$y_i$, with $i$ running over all site numbers. So in our model, any $y_i$ can be either zero or $1/2$, depending on whether the $i$th site is empty or occupied, respectively. Moreover, two consecutive $x$ labels are either equal or differ by $1/2$.

\begin{figure}[H]\centering
	\begin{tikzpicture}[baseline, scale=1, every node/.style={scale=1}]
		\draw[thick] (0,0)--(2.7,0);
		\draw[thick,dashed] (2.7,0)--(3.3,0);
		\draw[thick] (3.3,0)--(4,0);
		\draw[thick] (4,0)--(7.7,0);
		\draw[thick] (1,0)--(1,1);
		\draw[thick] (2,0)--(2,1);
		\draw[thick] (4,0)--(4,1);
		\draw[thick] (5,0)--(5,1);
		\draw[thick] (6,0)--(6,1);
		\draw[thick] (7,0)--(7,1);
		\draw[thick,dashed] (7.7,0)--(8.3,0);
		\draw[thick] (8.3,0)--(9,0);
		\draw[thick] (9,0)--(11,0);
		\draw[thick] (9,0)--(9,1);
		\draw[thick] (10,0)--(10,1);
		\node[below] at (.5,0) {$x_0$};
		\node[below] at (1.5,0) {$x_1$};
		\node[below] at (4.5,0) {$x_{i-1}$};
		\node[below] at (5.5,0) {$x_i$};
		\node[below] at (6.5,0) {$x_{i+1}$};
		\node[below] at (9.5,0) {$x_{l-1}$};
		\node[below] at (10.5,0) {$x_l$};
		\node[above] at (1,1) {$y_1$};
		\node[above] at (2,1) {$y_2$};
		\node[above] at (4,1) {$y_{i-1}$};
		\node[above] at (5,1) {$y_i$};
		\node[above] at (6,1) {$y_{i+1}$};
		\node[above] at (7,1) {$y_{i+2}$};
		\node[above] at (9,1) {$y_{l-1}$};
		\node[above] at (10,1) {$y_l$};
		\draw[thick, dotted] (2.5,0.5)--(3.5,0.5);
		\draw[thick, dotted] (7.5,0.5)--(8.5,0.5);
		\draw[blue, dashed] (5.5,.5) ellipse (1cm and 1.25cm);
	\end{tikzpicture}\caption{A typical fusion chain.}\label{fig:fusiontree}
\end{figure}
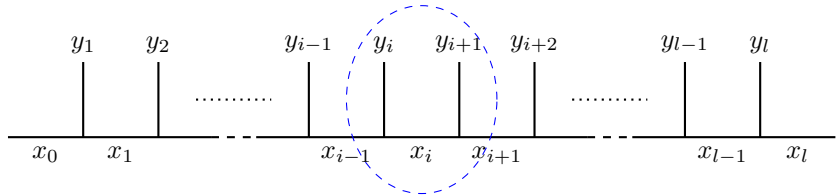

Consider the tree-like shape in Figure~\ref{fig:fusiontree} for a given configuration of $y$ labels. 

\begin{figure}[H]
\centering\begin{tikzpicture}[baseline, scale=1, every node/.style={scale=1}]
		\draw [thick] (0,0)--(3,0);
		\draw [thick] (1,0)--(1,1);
		\draw [thick] (2,0)--(2,1);
		\node [above] at (1,1) {\small$y_l$};
		\node [above] at (2,1) {\small$y_1$};
		\node [below] at (0.25,0) {\small$x_{l-1}$};
		\node [below] at (1.5,0) {\small$x_{l}=x_0$};
		\node [below] at (2.75,0) {\small$x_{1}$};
	\end{tikzpicture}\caption{The $l$th part of a closed chain.}\label{fig:closedtree}
\end{figure}
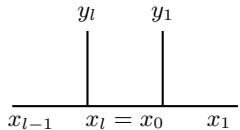

\noindent This tree is called a fusion \emph{chain} of size $l$ or with $l$ \emph{sites}, if the lower labels $x$---which are also assumed to be chosen from $\mathcal{L}_k$---are consistent with fusion rules, that is, if $x_j$ is a 
fusion channel of the fusion of $x_{j-1}$ with $y_j$, for all $1\leqslant j\leqslant l$. 
 The part indicated by the blue ellipse in Figure~\ref{fig:fusiontree} is called the $i$th part or, if we do not need to be explicit, a \emph{local} part of the fusion chain.
In this paper, we consider two types of chains---\emph{open} chains, for which both
labels $x_0$ and $x_l$ are fixed but arbitrary, and \emph{closed} chains, for which we impose the \emph{periodic} boundary condition $x_0=x_l$ on the fusion chains. In the latter case, Figure~\ref{fig:closedtree} shows what we call the $l$th part of a closed chain.

Using the notion of the fusion chain, we can now introduce the Hilbert space of our models, one corresponding to open chains and one corresponding to closed chains.  That $y$ labels in our models are restricted to take on either $0$ or $1/2$ makes it possible to introduce a short notation for fusion chains, since knowing only $x$ labels in a fusion chain, uniquely  determines the $y$ labels. In fact, if $x_{i-1}$ and $x_{i}$ differ by
$1/2$, then $y_i = 1/2$, and if $x_{i-1} = x_{i}$, then $y_i=0$. 
Therefore, the fusion chain in Figure~\ref{fig:fusiontree} can be simply indicated by the following ket:

\begin{align}\label{eq:ket}
	|x_0,x_1,\ldots,x_{i-1},x_i,x_{i+1},\ldots,x_{l-1},x_l\rangle.
\end{align}
For closed chains, we  can even simplify the notation further by dropping  the last
label $x_l = x_0$:
\begin{align}\label{eq:pket}
	|x_0,x_1,\ldots,x_{i-1},x_i,x_{i+1},\ldots,x_{l-1}\rangle.
\end{align}
We declare, for given $k$ and $l$, that the kets introduced in \eqref{eq:ket} and \eqref{eq:pket} constitute an orthonormal basis for the Hilbert spaces $\mathscr{H}_{\text{op}}(k,l)$ and $\mathscr{H}_{\text{cl}}(k,l)$ corresponding to open-chain and closed-chain models, respectively. The orthonormality of the basis-elements in each case is defined as
\begin{align}
	\langle x_0^{},x_1^{},\ldots,x_\nu^{} |x_0',x'_1,\ldots,x'_\nu\rangle=\prod_{i=0}^{\nu}\mathit{\delta}_{x_i^{}x'_i},
\end{align}
where $\nu$ is either $l-1$ or $l$, depending on whether the chain is closed or open, respectively.
\subsection{The Hamiltonian of the Model}
Here we introduce the Hamiltonians $H_{\text{op}}(k,l)$ and $H_{\text{cl}}(k,l)$ of the anyon models corresponding to open and closed chains, respectively. They act on Hilbert spaces $\mathscr{H}_{\text{op}}(k,l)$ and $\mathscr{H}_{\text{cl}}(k,l)$ introduced above,  correspondingly. In this section, we assume that $k$  and $l$ are fixed given numbers and do not write them explicitly.  Both Hamiltonians that we consider in this paper, which we  denote them simply by $H$, have the following form:
\begin{align}\label{eq:hamiltonian}
	H:=\sum_{i=1}^{\nu}h_i,
\end{align}
where $\nu$ is either $l-1$ or $l$, depending on whether it acts on closed or open chains, respectively. Each  $h_i$, which we call the $i$th \emph{local} Hamiltonian, is a sum of nine linear operators  in its own. Each one of these linear operators is defined to act non-trivially only on the $i$th part of the chain, in a way that is explained in detail in Subsection~\ref{subsec:localhamiltonian}. In the remainder of this section though, we describe the  general aspects of each one of these terms and fix sum notations.

As mentioned, in the models we consider, each $h_i$ is a sum of nine terms.
Four of these terms act diagonally by assigning energy according to whether
the sites $i$ and $i+1$ are occupied or not. Thus, these terms act as a chemical potential, if  both sites are not occupied simultaneously,  
and they act as an interaction term, if both sites are occupied simultaneously.
We denote these terms by
$h_{i,\mu_{00}}$,
$h_{i,\mu_{0\frac{1}{2}}}$,
$h_{i,\mu_{\frac{1}{2}0}}$, and
$h_{i,\mu_{\frac12\frac12}}$.

The fifth term we consider, is an interaction term $h_{i,J}$
between two neighboring anyons that acts in a way similar to the Heisenberg interaction
for spin-1/2 chains, that is, it assigns an energy that depends on the fusion channel of the
two neighboring anyons.  

The next two terms deal with hopping of the anyons. If an anyon is adjacent to an empty site, we consider the possibility of the anyon hopping to this empty site. The terms corresponding to this process are denoted by $h_{i,t}$ and $h'_{i,t}$  for hopping to right and left,
respectively. These types of terms were first considered in \cite{dilute}.

The last two terms, denoted by $h_{i,\mathit{\Delta}}$ and $h'_{i,\mathit{\Delta}}$, describe the creation and annihilation of two neighboring anyons. These are the new terms that we consider in
our model. In the presence of these terms, the number of anyons can fluctuate.

\subsubsection{Different Terms in the Local Hamiltonian $h_i$}\label{subsec:localhamiltonian}
To specify how the local Hamiltonian $h_i$ acts, consider a typical fusion chain, closed  or open, and focus on the $i$th part of this chain, which can be viewed as
\begin{align*}
\begin{tikzpicture}[baseline={([yshift=-1.9ex]current bounding box.center)},vertex/.style={anchor=base,
    circle,fill=black!25,minimum size=18pt,inner sep=2pt}]
        \draw[thick] (0,0)--(1.5,0);
		\draw[thick] (.5,0)--(.5,.5);
		\draw[thick] (1,0)--(1,.5);
		\node [above] at (.5,.5) {$m$};
		\node [above] at (1,.5) {$n$};
		\node [below] at (0.25,0) {$x$};
		\node [below] at (.75,0) {$y$};
		\node [below] at (1.25,0) {$z$};
\end{tikzpicture}\,:=|x,y,z\rangle,
\end{align*}
where $m$ and $n$ are either zero or $1/2$, and $y$ and $z$ are consistent with the fusion rules. 
We recall that the labels $m$ and $n$ are determined by the labels $x$, $y$, and $z$.

As mentioned in the pervious section, for each $i$, the local Hamiltonian $h_i$ is a sum of nine terms as follows: 
\begin{align}
	h_i=h_{i,\mu_{00}}+h_{i,\mu_{0\frac{1}{2}}}+h_{i,\mu_{\frac{1}{2}0}}+h_{i,\mu_{\frac12\frac12}}+h_{i,J}+h_{i,t}+h'_{i,t}+h_{i,\mathit{\Delta}}+h'_{i,\mathit{\Delta}},
\end{align}
where each term acts non-trivially only on the $i$th part of the chain and act as identity on other parts. Since, in this section, we always consider the $i$th part of the fusion chain and the $i$th local Hamiltonian, we suppress the subscript $i$ everywhere. In order to specify the model,
we now define how the different types of terms act on the local part of a fusion chain.

\vspace{.3cm}

\noindent\emph{The Diagonal Terms.}\hspace{.2cm}
The terms that assign an energy depending on the occupation of the neighboring sites, 
act diagonally. Explicitly, we write these terms as
\begin{align}
	h_{\mu_{00}}|x,x,x\rangle &:=\mu_{00}\,|x,x,x\rangle,\\
	h_{\mu_{0\frac{1}{2}}}|x,x,y\rangle &:=\mu_{0\frac{1}{2}}\,|x,x,y\rangle,\\
	h_{\mu_{\frac{1}{2}0}}|x,y,y\rangle &:=\mu_{\frac{1}{2}0}\,|x,y,y\rangle,\\
	h_{\mu_{\frac{1}{2}\frac{1}{2}}}|x,y,z\rangle &:=\mu_{\frac{1}{2}\frac{1}{2}}\,|x,y,z\rangle.
\end{align}
Recall that, if two neighboring labels in the kets above are different, their values differ by $1/2$. This
means, for instance, that $h_{\mu_{0\frac{1}{2}}}$ assigns an energy $\mu_{0\frac{1}{2}}$
if the first site is empty and the second site is occupied and, otherwise, this term acts by zero.
Similar considerations apply to the other terms.

We now turn our attention to the terms that act in a non-diagonal way and explain how they actually 
act in more detail.

\vspace{.3cm}

\noindent\emph{The Interaction Term.}\hspace{.2cm}
In our models, for two anyons sitting on neighboring sites,  we include an interaction term $h_J$  that
 assigns an energy which depends on the fusion channel of the two anyons. This interaction
term was introduced in the original paper \cite{feiguin}.
In particular, we demand for this term to assign an energy $J$, in the case two neighboring $1/2$ anyons fuse to zero, and to assign zero energy otherwise. Hence, in order to define $h_J$ we need to go to a basis for which the fusion channel of the two $1/2$ neighboring anyons is explicit.  To explain all of this, it is more illustrative to use the fusion  tree notation rather than the ket notation. This is illustrated in the following:

\begin{align}\label{eqn:interaction}
h_J\Bigg(\begin{tikzpicture}[baseline={([yshift=-3.87ex]current bounding box.center)},vertex/.style={anchor=base,
    circle,fill=black!25,minimum size=18pt,inner sep=2pt}]
        \draw[thick] (.15,0)--(1.35,0);
		\draw[thick] (.75,.5)--(.5,.75);
		\draw[thick] (.75,.5)--(1,.75);
		\draw[thick] (.75,0)--(.75,.5);
		\node [above] at (1,.75) {$\frac{1}{2}$};
		\node [above] at (.5,.75) {$\frac{1}{2}$};
		\node [below] at (0.4,0) {$x$};
		\node [right] at (.75,0.3) {$u$};
		\node [below] at (1.05,0) {$z$};
  \end{tikzpicture}\Bigg)
	:=J\,
	\,\delta_{xz}\delta_{u0}\begin{tikzpicture}[baseline={([yshift=-3.87ex]current bounding box.center)},vertex/.style={anchor=base,
    circle,fill=black!25,minimum size=18pt,inner sep=2pt}]
        \draw[thick] (.15,0)--(1.35,0);
		\draw[thick] (.75,.5)--(.5,.75);
		\draw[thick] (.75,.5)--(1,.75);
		\draw[thick] (.75,0)--(.75,.5);
		\node [above] at (1,.75) {$\frac{1}{2}$};
		\node [above] at (.5,.75) {$\frac{1}{2}$};
		\node [below] at (0.4,0) {$x$};
		\node [right] at (.75,0.3) {$u$};
		\node [below] at (1.05,0) {$z$};
  \end{tikzpicture}\,\raisepunct{.}
\end{align}
Here, as demanded, $\delta_{u0}$ takes care of assigning energy $J$ to the zero channel only, and $\delta_{xz}$ takes care of the consistency of the last fusion tree with the fusion rules. The term $h_J$ is defined to act by zero on any other configuration except the one mentioned above. 

To know how $h_J$ acts on a local fusion chain, we exploit Equations~\eqref{eq:fequation} and \eqref{eqn:interaction} and we come up with the following:
\begin{align}
	h_{J}\bigg(\begin{tikzpicture}[baseline={([yshift=-2.6ex]current bounding box.center)},vertex/.style={anchor=base,
    circle,fill=black!25,minimum size=18pt,inner sep=2pt}]
        \draw[thick] (0,0)--(1.5,0);
		\draw[thick] (.5,0)--(.5,.5);
		\draw[thick] (1,0)--(1,.5);
		\node [above] at (.5,.5) {$\frac{1}{2}$};
		\node [above] at (1,.5) {$\frac{1}{2}$};
		\node [below] at (0.25,0) {$x$};
		\node [below] at (.75,0) {$y$};
		\node [below] at (1.25,0) {$z$};
  \end{tikzpicture}\bigg)&=J\,\sum_{u}F^{x\frac{1}{2}\frac{1}{2}}_{z;\,yu}\,\begin{tikzpicture}[baseline={([yshift=-3.7ex]current bounding box.center)},vertex/.style={anchor=base,
    circle,fill=black!25,minimum size=18pt,inner sep=2pt}]
        \draw[thick] (.15,0)--(1.35,0);
		\draw[thick] (.75,.5)--(.5,.75);
		\draw[thick] (.75,.5)--(1,.75);
		\draw[thick] (.75,0)--(.75,.5);
		\node [above] at (1,.75) {$\frac{1}{2}$};
		\node [above] at (.5,.75) {$\frac{1}{2}$};
		\node [below] at (0.4,0) {$x$};
		\node [right] at (.75,0.3) {$u$};
		\node [below] at (1.05,0) {$z$};
  \end{tikzpicture}\,\delta_{u0}=
  J\,F^{x\frac{1}{2}\frac{1}{2}}_{z;\,y0}\,\begin{tikzpicture}[baseline={([yshift=-3.87ex]current bounding box.center)},vertex/.style={anchor=base,
    circle,fill=black!25,minimum size=18pt,inner sep=2pt}]
        \draw[thick] (.15,0)--(1.35,0);
		\draw[thick] (.75,.5)--(.5,.75);
		\draw[thick] (.75,.5)--(1,.75);
		\draw[thick] (.75,0)--(.75,.5);
		\node [above] at (1,.75) {$\frac{1}{2}$};
		\node [above] at (.5,.75) {$\frac{1}{2}$};
		\node [below] at (0.4,0) {$x$};
		\node [right] at (.75,0.3) {$0$};
		\node [below] at (1.05,0) {$z$};
  \end{tikzpicture}\,\cdot
\end{align}
Switching back to the original basis by employing the inverse of an $F$-move and using the fact that the $F$-matrices we use are their own inverses, one gets
\begin{align}
	h_{J}\bigg(\begin{tikzpicture}[scale=1,baseline=-.5ex]
        \draw[thick] (0,0)--(1.5,0);
		\draw[thick] (.5,0)--(.5,.5);
		\draw[thick] (1,0)--(1,.5);
		\node [above] at (.5,.5) {$\frac{1}{2}$};
		\node [above] at (1,.5) {$\frac{1}{2}$};
		\node [below] at (0.25,0) {$x$};
		\node [below] at (.75,0) {$y$};
		\node [below] at (1.25,0) {$z$};
  \end{tikzpicture}\bigg)=J \delta_{xz}\,\sum_{v}\bigg(F^{x\frac{1}{2}\frac{1}{2}}_{z;\,y0}\,F^{x\frac{1}{2}\frac{1}{2}}_{z;\,0v}\,\begin{tikzpicture}[scale=1,baseline=-.5ex]
        \draw[thick] (0,0)--(1.5,0);
		\draw[thick] (.5,0)--(.5,.5);
		\draw[thick] (1,0)--(1,.5);
		\node [above] at (.5,.5) {$\frac{1}{2}$};
		\node [above] at (1,.5) {$\frac{1}{2}$};
		\node [below] at (0.25,0) {$x$};
		\node [below] at (.75,0) {$v$};
		\node [below] at (1.25,0) {$z$};
  \end{tikzpicture}\bigg)\raisepunct{.}
\end{align}
The operator $h_J$ acts by zero on any other configuration of the local fusion chain other than the ones mentioned above. Plugging the $F$-symbols introduced in Appendix~\ref{app:f-symbols} into the equation above, we have:
\begin{align}\label{eq:interaction}
	h_{J}\bigg(\begin{tikzpicture}[scale=1,baseline=-.5ex]
        \draw[thick] (0,0)--(1.5,0);
		\draw[thick] (.5,0)--(.5,.5);
		\draw[thick] (1,0)--(1,.5);
		\node [above] at (.5,.5) {$\frac{1}{2}$};
		\node [above] at (1,.5) {$\frac{1}{2}$};
		\node [below] at (0.25,0) {$x$};
		\node [below] at (.75,0) {$y$};
		\node [below] at (1.25,0) {$z$};
  \end{tikzpicture}\bigg)=J \delta_{xz}\,\sum_{v}\bigg(\frac{\sqrt{d_yd_v}}{d_xd_{1/2}}\,\begin{tikzpicture}[scale=1,baseline=-.5ex]
        \draw[thick] (0,0)--(1.5,0);
		\draw[thick] (.5,0)--(.5,.5);
		\draw[thick] (1,0)--(1,.5);
		\node [above] at (.5,.5) {$\frac{1}{2}$};
		\node [above] at (1,.5) {$\frac{1}{2}$};
		\node [below] at (0.25,0) {$x$};
		\node [below] at (.75,0) {$v$};
		\node [below] at (1.25,0) {$z$};
  \end{tikzpicture}\bigg)\raisepunct{.}
\end{align}
Of course, considering fusion rules, the sum above has at most two terms in our model.\\

\noindent\emph{The Hopping Terms.}\hspace{.2cm}
The models we consider allow for the possibility for a $1/2$ anyon to hop onto a neighboring
site, provided this site is empty. We denote the strength of the hopping process by $t$.
This hopping process was first considered in\cite{feiguin}.
Explicitly, the hopping terms act as 
\begin{align}
	h_{t}|x,y,y\rangle &:=t\,|x,x,y\rangle,\\
	h'_{t}|x,x,y\rangle &:=t\,|x,y,y\rangle.
\end{align} 
Here, again, different letters in the kets refer to different labels and for all other ket configurations,
$h_t$ and $h'_t$ are defined to act by zero.\\
\\
\emph{The Creation and Annihilation Terms.}\hspace{.2cm}
Finally, we introduce the terms that allow for the number of anyons to fluctuate. This is achieved
by  considering the possibility for a process in which a pair of $1/2$ anyons is created out of the vacuum on two neighboring empty sites as well as the possibility for a process in which a pair of $1/2$ anyons sitting on neighboring sites are annihilated. We assume the same strength $\mathit{\Delta}$ for both of these  processes to guarantee the Hermiticity of the  Hamiltonian. 

The procedure of defining the creation term is along the lines that we had for interaction term. Hence, making  explanations short, we have: 
\begin{align}
	h_{\mathit{\Delta}}\bigg(\begin{tikzpicture}[scale=1,baseline=-.5ex]
        \draw[thick] (0,0)--(1.5,0);
		\draw[thick] (.5,0)--(.5,.5);
		\draw[thick] (1,0)--(1,.5);
		\node [above] at (.5,.5) {$0$};
		\node [above] at (1,.5) {$0$};
		\node [below] at (0.25,0) {$x$};
		\node [below] at (.75,0) {$x$};
		\node [below] at (1.25,0) {$x$};
  \end{tikzpicture}\bigg)&:=\mathit{\Delta}\,
\begin{tikzpicture}[scale=1,baseline=-.5ex]
        \draw[thick] (.15,0)--(1.35,0);
		\draw[thick] (.75,.5)--(.5,.75);
		\draw[thick] (.75,.5)--(1,.75);
		\draw[thick] (.75,0)--(.75,.5);
		\node [above] at (1,.75) {$\frac{1}{2}$};
		\node [above] at (.5,.75) {$\frac{1}{2}$};
		\node [below] at (0.4,0) {$x$};
		\node [right] at (.75,0.3) {$0$};
		\node [below] at (1.05,0) {$x$};
  \end{tikzpicture}=
  \mathit{\Delta}\,\sum_{u}F^{x\frac{1}{2}\frac{1}{2}}_{x;\,0u}\,
  \begin{tikzpicture}[scale=1,baseline=-.5ex]
        \draw[thick] (0,0)--(1.5,0);
		\draw[thick] (.5,0)--(.5,.5);
		\draw[thick] (1,0)--(1,.5);
		\node [above] at (.5,.5) {$\frac{1}{2}$};
		\node [above] at (1,.5) {$\frac{1}{2}$};
		\node [below] at (0.25,0) {$x$};
		\node [below] at (.75,0) {$u$};
		\node [below] at (1.25,0) {$x$};
  \end{tikzpicture}\,\raisepunct{.}
\end{align}
In the first step, we create a pair of $1/2$ anyons out of the vacuum so that they are
in the fusion channel zero. In the second step, we rewrite the obtained configuration
in the basis we use to describe the Hamiltonians.
Plugging the explicit form of the $F$-symbols, yields:
\begin{align}\label{eq:creation}
	h_{\mathit{\Delta}}\bigg(\begin{tikzpicture}[scale=1,baseline=-.5ex]
        \draw[thick] (0,0)--(1.5,0);
		\draw[thick] (.5,0)--(.5,.5);
		\draw[thick] (1,0)--(1,.5);
		\node [above] at (.5,.5) {$0$};
		\node [above] at (1,.5) {$0$};
		\node [below] at (0.25,0) {$x$};
		\node [below] at (.75,0) {$x$};
		\node [below] at (1.25,0) {$x$};
  \end{tikzpicture}\bigg)=\mathit{\mathit{\Delta}}\,\sum_{u}\bigg(\sqrt{\frac{d_u}{d_xd_{1/2}}}\,\begin{tikzpicture}[scale=1,baseline=-.5ex]
        \draw[thick] (0,0)--(1.5,0);
		\draw[thick] (.5,0)--(.5,.5);
		\draw[thick] (1,0)--(1,.5);
		\node [above] at (.5,.5) {$\frac{1}{2}$};
		\node [above] at (1,.5) {$\frac{1}{2}$};
		\node [below] at (0.25,0) {$x$};
		\node [below] at (.75,0) {$u$};
		\node [below] at (1.25,0) {$x$};
  \end{tikzpicture}\bigg)\raisepunct{.}
\end{align}
Here again the sum above consists of at most two terms in our models.\\

The process conjugate to the creation of two anyons on neighboring sites is the
annihilation of two anyons sitting on neighboring sites. The latter is possible only if the fusion
channel of the two anyons is zero. 
Using similar arguments as for the creation term, we get:
\begin{align}\label{eq:annihilation}
	h'_{\mathit{\Delta}}\bigg(\begin{tikzpicture}[scale=1,baseline=-.5ex]
        \draw[thick] (0,0)--(1.5,0);
		\draw[thick] (.5,0)--(.5,.5);
		\draw[thick] (1,0)--(1,.5);
		\node [above] at (.5,.5) {$\frac{1}{2}$};
		\node [above] at (1,.5) {$\frac{1}{2}$};
		\node [below] at (0.25,0) {$x$};
		\node [below] at (.75,0) {$y$};
		\node [below] at (1.25,0) {$z$};
  \end{tikzpicture}\bigg)=\delta_{xz}\mathit{\mathit{\Delta}}\,\sqrt{\frac{d_y}{d_xd_{1/2}}}\,\begin{tikzpicture}[scale=1,baseline=-.5ex]
        \draw[thick] (0,0)--(1.5,0);
		\draw[thick] (.5,0)--(.5,.5);
		\draw[thick] (1,0)--(1,.5);
		\node [above] at (.5,.5) {$0$};
		\node [above] at (1,.5) {$0$};
		\node [below] at (0.25,0) {$x$};
		\node [below] at (.75,0) {$x$};
		\node [below] at (1.25,0) {$x$};
  \end{tikzpicture}\,\raisepunct{.}
\end{align}

It is straightforward to see that each $h_i$ is a symmetric operator and, consequently, $H$ is symmetric (in fact, Hermitian) and a physically acceptable Hamiltonian.

\section{Ground States of the Model Hamiltonians}
\label{sec:exact-groundstates}

In this section, we start with the analysis of the dilute anyon model we introduced in the
previous section. Since the number of parameters in the model is rather large and also the
size of the Hilbert space grows quickly with both $k$ and system size $l$, we limit ourselves to
special choices for the parameters and analyze those analytically, supported by
numerical calculations. 

A possible strategy, which has already been proven valuable in similar models, is to choose values for the parameters to make each local Hamiltonian $h_i$ a projector, that is,
$h_i^2 = h_i$. Consequently, the eigenvalues of each $h_i$ is either zero or one.
The efficiency of this strategy is due to the fact that, if the local Hamiltonians $h_i$ are projectors, then a state $|\psi\rangle$ is the \emph{zero}-energy ground state of the total Hamiltonian~\eqref{eq:hamiltonian} if and only if $h_i|\psi\rangle=0$ for all $i$ or, equivalently, if and only if	
\begin{align}\label{eqn:kernel}
	\ker(H)=\bigcap_{i=1}^{\nu}\ker(h_i),
\end{align}
for the kernel of these operators. In some cases, it is possible to find all the states $|\psi\rangle$ such that $h_i|\psi\rangle=0$ for all $i$.
Two famous examples for which this has been done are the spin-1/2
Majumdar-Ghosh model~\cite{mga,mgb}, in which the projector projects onto the spin-3/2 states of
three neighboring spin-1/2's, and the spin-1 AKLT model \cite{AKLT1,AKLT2}, in which the projector
projects onto the spin-2 states of two neighboring spin-1's.

Typically, if the Hamiltonian is a sum of projectors and one can find zero-energy ground states,
the system is gapped. This has been proven in some cases. Because of the gap, these ground states typically describe the physics of the corresponding models even if one perturbs away from the point where the Hamiltonian is a sum of projectors. We note that it can of course happen that the ground state does not have zero energy. In these cases, one has to resort to other techniques to analyze the model.  

Figuring out the values for the parameters to turn the local Hamiltonians into projectors is rather straightforward in our case. This stems from the fact that when a local Hamiltonian acts on the corresponding local piece  $| x_{i-1}, x_i, x_{i+1}\rangle$ of the fusion chain, 
it does not alter the outer labels $x_{i-1}$ and $x_{i+1}$. Therefore, we can consider the
local subspaces $\{ |x,x,x\pm1/2 \rangle, |x,x\pm1/2,x\pm1/2 \rangle \}$ and
$\{ |x,x,x\rangle , |x,x-1/2,x\rangle , |x,x+1/2,x\rangle \}$ separately. We note that the
states $|x,x-1/2,x-1\rangle$ and $|x,x+1/2,x+1\rangle$ form one-dimensional subspaces
in their own.

The constraints on the parameters of the Hamiltonian can straightforwardly be solved. 
Moreover, in this paper we are interested in a non-zero value for the parameters $J$, $t$, 
and $\mathit{\Delta}$. Out of the possible solutions, we pick the following set of values:
\begin{align}\label{eqn:parametervalues}
	\mu_{00}=\mu_{0\frac12}=\mu_{\frac{1}{2}0}=t=\frac{1}{2}\;\raisepunct{,}\; \mu_{\frac12\frac12}=0\;\raisepunct{,}\; J=\frac{1}{2}\;\raisepunct{,}\;\mathit{\Delta}=\frac{1}{2}\raisepunct{,}
\end{align}
which by assigning the same value $1/2$ to parameters $\mu_{00}$, $\mu_{0\frac{1}{2}}$, $\mu_{\frac{1}{2}0}$, and $t$, makes the model easier to handle.

Assigning these values to the parameters, we found that, for \emph{odd} values of $k$, there are indeed zero-energy ground states.
In the case of an open chain, we found that there are $(k+1)(k+2)(k+3)/6$ zero-energy ground states, provided that the chain is sufficiently long, $l\geqslant k$. For a closed chain, there are $(k+1)/2$ zero-energy ground states, if $l\geqslant k+1$. In each case, for system sizes smaller than these thresholds, the number of zero-energy ground states depends on the system size $l$ and decreases by increasing $l$ until it saturates at the numbers mentioned above. At this point of the parameter space and for odd values of $k$, we were able to determine an explicit closed form for these zero-energy ground states. This is discussed separately for the open and closed cases in
Subsections~\ref{subsec:ground states open} and \ref{subsec:ground states closed}.

We should note that in the case of \emph{even} $k$, it turns out that, if the system size is large enough,  there are no zero-energy ground states. The reason for the difference in behavior of the model for
$k$ even and odd, lies in the fact that the structure of the fusion rules is different for these
cases. As an example of this difference, we note that one has the fusion rule
$j \otimes k/2 = k/2 - j$. Thus, if $k$ is even, there is
an anyon type, namely $j=k/4$, such that $j \otimes k/2 = j$, while this is not the case when $k$ is odd.


\subsection{Zero-Energy Ground States, Open Chain}\label{subsec:ground states open}

In this subsection we first outline the general strategy that we followed to determine the zero-energy ground states for the described model Hamiltonian in the case of a given odd $k$ and a sufficiently large open chain with the parameter values~\eqref{eqn:parametervalues}. For an open chain with $l$ sites, the
Hamiltonian is given by $H = \sum_{i=1}^{l-1} h_i$. This is a sum over $l-1$ terms, where the
first acts on the first three labels $x_0$, $x_1$, and $x_2$, while the
last term acts on the last three labels  $x_{l-2}$, $x_{l-1}$, and $x_{l}$ of every given basis state $| x_0,x_1,\ldots,x_{l}\rangle$.

We start by considering the explicit form of these states for the cases $k=1$ and $k=3$, and explain the number of ground states for a given boundary condition. As becomes clear, the structure of the coefficients of the ground states are fairly simple for these special cases. Having understood the structure of  ground states for $k=1$ and $k=3$, one can generalize and find the ground states for general odd $k$, which we present at the end of this section.

We decompose $\mathscr{H}_{\text{op}}$, the Hilbert space introduced in Subsection~\ref{subsec:hilbert space} for an open chain, into $(k+1)^2$ disjoint sectors with fixed boundary-labels as follows:
\begin{align}
	\mathscr{H}_{\text{op}}=\bigoplus_{\mathclap{a,b\in \mathcal{L}_k}}\mathscr{H}_{\text{op}}^{ab},
\end{align}
where $\mathscr{H}_{\text{op}}^{ab}$ denotes the subspace of $\mathscr{H}_{\text{op}}$ spanned by those basis fusion kets $| x_0,x_1,\ldots,x_{l}\rangle$ in which $x_0=a$ and $x_l=b$. The Hamiltonian does not alter the values of
$x_0$ and $x_l$ when it acts on the corresponding ket, so we can search for zero-energy ground states in each subspace
$\mathscr{H}_{\text{op}}^{ab}$ separately. Below, when we write $H$ and $h_i$, we actually mean their
restrictions to $\mathscr{H}_{\text{op}}^{ab}$.

To describe the general strategy, let $|\psi^{ab}\rangle$ be a generic element of $\mathscr{H}_{\text{op}}^{ab}$, namely, 
\begin{align}\label{eq:ground state ij}
	|\psi^{ab}\rangle=\sum_{\{x_i\}}C_{\{x_i\}}^{ab}\,|a,x_{1},\ldots,x_{l-1},b\rangle,
\end{align}
where $C_{\{x_i\}}^{ab}$'s are, in general, complex numbers, $\{|a,x_{1},\ldots,x_{l-1},b\rangle\}_{\{x_i\}}$ is the basis of $\mathscr{H}_{\text{op}}^{ab}$  composed of fusion chains starting with label $a$ and ending with label $b$, and the sum runs through all possible intermediate labels $x_i$.
For $|\psi^{ab}\rangle$ to be a zero-energy ground state, it must be in the kernel of the Hamiltonian $H$.
Consequently, by Equation~\eqref{eqn:kernel},  it must reside in the kernel of all local Hamiltonians $h_i$.
To determine a set of appropriate coefficients $C^{ab}_{\{x_i\}}$ for the zero-energy ground state(s),
one can act as follows. We first consider the relation $h_1|\psi^{ab}\rangle=0$, which gives
rise to some constraints on the coefficients $C_{\{x_i\}}^{ab}$. Typically, it relates several
coefficients. We denote the generic state satisfying these relations $|\psi_1^{ab}\rangle$.
One continues by considering that $h_2|\psi_1^{ab}\rangle=0$, which gives rise to more
constraints among the coefficients. Continuing this way, one finds that the number of states, 
satisfying the increasing number of constraints, decreases until it eventually saturates.

We start by looking at $k=1$ in which there are four different boundary conditions. By following
the procedure outlined above, one can show explicitly that for each boundary condition, there is one
zero-energy ground state. In addition, one finds that the coefficients describing these ground states
take a simple form, namely $C_{\{x_i\}}^{ab}=(-1)^{\#(1/2)}$, where $\#(1/2)$ is the number of
$1/2$ labels present in the corresponding basis state $|a,x_{1},\ldots,x_{l-1},b\rangle$.
The ground states thus take the following simple form:
\begin{align}
\label{eq:ground state k=1}
|\psi^{ab} \rangle=\sum_{\{x_i\}} (-1)^{\#(1/2)}
|a,x_{1},\ldots,x_{l-1},b\rangle,
\end{align}
with the sum over all states in $\mathscr{H}_{\text{op}}^{ab}$.
Because the coefficients appeared in all of these four ground states obey the same rule, we say that these states are all of the same type. As become clear shortly, for higher values of
$k$, there are ground states of different type.
We can summarize the number of ground states for the different boundary conditions in terms
of the following matrix: 
\begin{align}\label{eqn:matrixk=1}
M^{(k=1)} =
\begin{bmatrix}
1 & 1\\
1 & 1
\end{bmatrix},
\end{align}
where the rows and columns correspond to the labels $0$ and $1/2$.

The case $k=3$ is more complicated than $k=1$. There are sixteen different boundary conditions 
and the constraints imposed by the Hamiltonian are more complicated. We used the same line of
arguments as for $k=1$, but we guided ourselves by diagonalizing the Hamiltonian for small system
sizes. For system size $l\geqslant 3$, there is a total of twenty zero-energy ground states. 
We observe that all these twenty ground states fall into two main types, as compared to only one type for $k=1$ case. There are ground states with the property that in their expansions in terms of basis states, all basis states contribute. In other words, the labels in $|a,x_{1},\ldots,x_{l-1},b\rangle$ are taken from the set $\{0,1/2,1,3/2\}$. We call these, ground states of \emph{type one}.
The coefficients of this type of ground states turns out to take the following pattern:
\begin{align}
	C_{\{x_i\}}^{ab}=(-1)^{[\#(1/2)+\#(3/2)]}_{\phantom{1/2}}\,d_{1/2}^{-3/2\times\#(1,1/2)},
\end{align}
where $\#(1/2)$ and  $\#(3/2)$ are the number of labels in $|a,x_{1},\ldots,x_{l-1},b\rangle$ that are $1/2$ and $3/2$, respectively, and  $\#(1,1/2)$ is the number of
ordered pairs $(x_i,x_{i+1})$ in this ket that are equal to $(1,1/2)$---taking the cases $i=0$ and $i=l-1$ into account  as well. We remind that $d_{1/2}$ is the quantum dimension
of the anyon type $1/2$, which for $k=3$ is equal to the golden ratio $\phi:=(1+\sqrt{5})/2$, by Equation~\eqref{eq:ds}.

There is another type of ground states, which we call ground states of \emph{type two}, such that, in their expansions, only those basis states $|a,x_{1},\ldots,x_{l-1},b\rangle$ contribute that have their labels in the set $\{1/2,1\}$. The coefficients in this case, take the following pattern:
\begin{align}
	C_{\{x_i\}}^{ab}=(-1)_{\phantom{1/2}}^{\#(1/2)}\, d_{1/2}^{\frac{1}{2}\times\#(1,1/2)}.
\end{align}

From the structure of the two types of ground states, we can deduce the number of zero-energy
ground states for each boundary condition. The ground state of the first type is present for
all sixteen boundary conditions, while the ground state of the second type only occurs if both labels $a$ and $b$ belong to the set $\{1/2,1\}$. We indicate this by the matrix
\begin{align}\label{eqn:matrixk=3}
M^{(k=3)} =
\begin{bmatrix}
1 & 1 & 1 & 1\\
1 & 2 & 2 & 1\\
1 & 2 & 2 & 1\\
1 & 1 & 1 & 1
\end{bmatrix}
\raisepunct{,}
\end{align}
where the rows and columns correspond to the labels $0$, $1/2$, $1$, and $3/2$ in this order.

Having understood the structure of the zero-energy ground states for $k=1$ and $k=3$, we
now consider a generic odd $k$. Guided by numerical diagonalization of small system sizes,
we found the ground states for odd $k$ in general. For large enough system size, namely for
$l\geqslant k$, the number of zero-energy ground states of the system is $(k+1)(k+2)(k+3)/6$ 
that can be viewed as $(k+1)/2$ \emph{different types}, according to the labels present in the basis states appearing in their expansions.

For a  ground state of \emph{type one}, the labels $x_i$ of the basis states  $|x_{0},x_{1},\ldots,x_{l}\rangle$ with non-zero coefficients
all belong to the  set $\{0,1/2,1,\ldots,(k-2)/2,(k-1)/2,k/2\}$. In other words, all basis states have non-zero
coefficients in this case. For a ground state of \emph{type two}, all the labels $x_i$ belong to the set
\mbox{$\{1/2,1,\ldots,(k-2)/2,(k-1)/2\}$}, that is, any basis state with at least one label $x_i$ equal to zero or $k/2$  has
zero coefficient. In general, for a state of \emph{type} $n$, the labels of basis states with non-zero coefficient belong to the following set:
\begin{align}\label{eqn:labelset}
 \Big\{\frac{n-1}{2}\raisepunct{,}\frac{n+1}{2}\raisepunct{,}\cdots\raisepunct{,}\frac{k-(n+1)}{2}\raisepunct{,}\frac{k-(n-1)}{2}\Big\}\raisepunct{.}	
\end{align}

To explicitly express the coefficients of the basis states in each type of these ground states, we introduce a piece of  notation first. For fixed odd $k$ and integer $n$, $1\leqslant n \leqslant (k+1)/2$, and for $i=n,n+1,\ldots,(k+1)/2$,  we define $D(k,n,i)$ by
\begin{align}\label{eq:Ds}
	D(k,n,i)    =\left\{\hspace{-.2cm}\begin{array}{rl}
		d_{1/2}^{1/2}\,d_{(k-1)/4}^{-1}\,d_{(i-1)/2}^{-1/2}\,d_{i/2}^{-1/2}\,d_{(i-n-1)/4}^{\phantom{-1/2}}\,d_{(k-n-i)/4}^{\phantom{-1/2}},&\text{if $i-n$ is odd,}\\\\
		d_{1/2}^{1/2}\,d_{(k-1)/4}^{-1}\,d_{(i-1)/2}^{-1/2}\,d_{i/2}^{-1/2}\,d_{(n+i-2)/4}^{\phantom{-1/2}}\,d_{(k-n+i+1)/4}^{\phantom{-1/2}},&\text{if $i-n$ is even.}
	\end{array}\right.
\end{align}
Now let $|\psi\rangle$ be a type-$n$ ground state and let $|x_{0},x_{1},\ldots,x_{l}\rangle$ be a basis state that appears in the expansion of $|\psi\rangle$ with constraints on $x_i$'s explained above. The coefficient of this basis state in the expansion of $|\psi\rangle$ is
\begin{align}\label{eqn:coefficients}
	(-1)^m\prod_{i=n}^{\mathclap{(k+1)/2}}[D(k,n,i)]^{\theta_i},
\end{align}
where $m$ is the number of half-integers in $|x_{0},x_{1},\ldots,x_{l}\rangle$,
\begin{align}\label{eqn:thetas}
	\theta_{i}:=\#\Big(\frac{i}{2}\raisepunct{,} \frac{i-1}{2}\Big)+\#\Big(\frac{k-i+1}{2}\raisepunct{,} \frac{k-i}{2}\Big)\raisepunct{,} \qquad\big(n\leqslant i\leqslant (k-1)/2\big)\raisepunct{,} 
\end{align}
and
\begin{align}\label{eqn:theta}
	\theta_{(k+1)/2}:=\#\Big(\frac{k+1}{4}\raisepunct{,} \frac{k-1}{4}\Big)\raisepunct{.} 
\end{align}
Here $\#(r,s)$ refers to the number of pairs $(x_i,x_{i+1})$ composed of labels in $|x_{0},x_{1},\ldots,x_{l}\rangle$ that are equal to the ordered pair $(r,s)$. This completes the description of the explicit form of the zero-energy
ground states.

We mention that the number of zero-energy ground states follows the same pattern as for
$k=1$ and $k=3$. Explicitly, the entries $m_{ij}$ of the matrix $M^{(k)}$, analog 
 to matrices~\eqref{eqn:matrixk=1} and \eqref{eqn:matrixk=3}, encoding the number of ground
 states are given by
 \begin{align}
 	m_{ij}=\frac{1}{2}\,\big[k+2-\max\{|k-4i|,|k-4j|\}\big]\raisepunct{,}
 \end{align}
 where $i$ and $j$ run over the values $0,1/2,\ldots,k/2$.

\subsection{Zero-Energy Ground States, Closed Chain}\label{subsec:ground states closed}

In this section, we deal with the ground states of the closed chain. Interestingly, we find that the number of zero-energy ground states differs
from the open case. As in the previous section, $k$ has to be odd, otherwise, no zero-energy
ground states exist, if the system is large enough.

For a closed chain, we have that $x_l = x_0$, therefore, as mentioned earlier, we
label the states in the Hilbert space by the kets
$|x_{0},x_{1},\ldots,x_{l-1}\rangle$. In terms of the Hilbert space of the open chain,
we have \begin{align}
	\mathscr{H}_{\text{cl}}=\bigoplus_{a\in\mathcal{L}_k}\mathscr{H}_{\text{op}}^{aa}.
\end{align}
The Hamiltonian is now a sum over $l$ terms $H = \sum_{i=1}^{l} h_i$---including one more
term compared to the Hamiltonian for the open case. This additional term acts on the label
$x_0 = x_l$, which is not left invariant by the Hamiltonian anymore.

To find the zero-energy ground state $|\psi\rangle$ of the closed chain, we take all the ground states of
the open chain with the boundary conditions $x_0 = x_l = a$. On these states, we need
to impose an additional constraint, namely, $h_l | \psi \rangle = 0$. This additional
constraint reduces the number of zero-energy ground states. The result is that there is exactly
one ground state for each `type' of ground state that was introduced in the previous
section. In this case, the unique zero-energy ground state of type $n$, denoted by
$|\text{GS}_n\rangle$, can be explicitly written as
\begin{align}
	|\text{GS}_n\rangle=\mathlarger{\sum}_{\{x_i\}}\bigg((-1)^m\prod_{i=n}^{\mathclap{(k+1)/2}}[D(k,n,i)]^{\theta_i}\bigg)|x_{0},x_{1},\ldots,x_{l-1}\rangle\raisepunct{,}
\end{align}
where the sum is over all possible labels chosen from the set \eqref{eqn:labelset}. As is apparent in the formula above, the form of the coefficients are exactly the same as for the open case, but this time $m$ refers to the number of half integers in $|x_{0},x_{1},\ldots,x_{l-1}\rangle$, and although $\theta_i$, $i=n,n+1,\ldots,(k+1)/2$, is defined as mentioned in Equations~\eqref{eqn:thetas} and \eqref{eqn:theta}, but this time $\#(r,s)$ refers to the number of ordered pairs $(x_i,x_{i+1})$ in $|x_{0},x_{1},\ldots,x_{l-1}\rangle$  equal to $(r,s)$, where we use the identification $x_l = x_0$.


\section{Integrability of the Model}
\label{sec:integrability}

In the previous section, we studied the model for the parameters chosen in
such a way that the Hamiltonian becomes a sum of projectors. This allowed us,
for odd $k$, to find the zero-energy ground states of the model. In this section,
we investigate if the parameters in the model can be chosen such that the
model becomes integrable. We refer to \cite{book:baxter} for an introduction on
the Yang--Baxter equation and transfer matrices. 

We start by briefly recalling the situation for the dense anyon chain where all $l$ sites are occupied, that is, $y_i = 1/2$ for all $i$. In this case, the
only types of terms that survive in the Hamiltonian are the `interaction' terms $h_{J}$ and $h_{\mu_{\frac{1}{2}\frac{1}{2}}}$.  Letting $J=\mu_{\frac{1}{2}\frac{1}{2}}=1$, $h_{\mu_{\frac{1}{2}\frac{1}{2}}}$ acts as the identity operator and $h_{i,J}$ acts as a projector that assigns an energy
$+1$ to two neighboring anyons, if they fuse to  the zero channel,
and assigns energy zero, if they fuse to  the one channel. Therefore, up to an overall shift, 
 $H = \sum_{i} h_{i,J=1}$. It is straightforward to see that the operators $e_i$, $i=1,2,\ldots,l$,  defined by $e_{i} = d_{1/2}\, h_{i,J=1}$ satisfy the Temperley--Lieb algebra, namely,
\begin{align}\label{eqn:TLalgebra}
\begin{array}{rll}
e_i^2 = d^{\phantom{2}}_{1/2}\, e_i^{\phantom{2}},&&\text{for all $i$}, \\
e_{i} e_{i\pm1} e_{i} = e_i,&&\text{for all $i$}, \\
{[}e_i,e_j{]}=0,&& \text{for $|i-j| \geqslant 2.$}	\\
\end{array}
\end{align}
It is instructive to give a pictorial representation of the Temperley--Lieb algebra.
Consider a chain of $l$ sites, corresponding to the $l$ spin-1/2 anyons of
the dense anyon model and, for each site, draw a vertical line as is depicted on the left in Figure~\ref{fig:TLoperators}. This picture 
is associated with the identity operator acting on the $l$ sites. The picture associated with the operator $e_i$ is depicted on the right panel of Figure~\ref{fig:TLoperators}. 
The operator $e_i$ is represented in a similar way, except that now sites $i$ and $i+1$ are
connected by a line at the top and a line at the bottom. 
\begin{figure}[H]\centering
\label{fig:TLoperators}
\begin{align*}
\bs{1}=\begin{tikzpicture}[scale=1.2,baseline=.38cm]
		\draw[thick] (0,0)--(0,.8);
		\draw[thick] (.6,0)--(.6,.8);
		\draw[thick] (1.2,0)--(1.2,.8);
		\draw[thick] (3,0)--(3,.8);
		\draw[thick] (3.6,0)--(3.6,.8);
		\draw[thick] (4.2,0)--(4.2,.8);
		\draw[thick, dotted] (.7,.4)--(1.1,.4);
		\draw[thick, dotted] (3.1,.4)--(3.5,.4);
		\draw [thick] (2.4,0)--(2.4,.8);
		\draw [thick] (1.8,0)--(1.8,.8);
		\node[below] at (0,0) {\small{$1$}};
		\node[below] at (.6,0) {\small{$2$}};
		\node[below] at (1.8,0) {\small{$i$}};
		\node[below] at (2.4,0) {\small{$i+1$}};
		\node[below] at (3.6,0) {\small{$l-1$}};
		\node[below] at (4.2,0) {\small{$l$}};
\end{tikzpicture},\quad
	e_i=\begin{tikzpicture}[scale=1.2,baseline=.38cm]
		\draw[thick] (0,0)--(0,.8);
		\draw[thick] (.6,0)--(.6,.8);
		\draw[thick] (1.2,0)--(1.2,.8);
		\draw[thick] (3,0)--(3,.8);
		\draw[thick] (3.6,0)--(3.6,.8);
		\draw[thick] (4.2,0)--(4.2,.8);
		\draw[thick, dotted] (.7,.4)--(1.1,.4);
		\draw[thick, dotted] (3.1,.4)--(3.5,.4);
		\node[below] at (0,0) {\small{$1$}};
		\node[below] at (.6,0) {\small{$2$}};
		\node[below] at (1.8,0) {\small{$i$}};
		\node[below] at (2.4,0) {\small{$i+1$}};
		\node[below] at (3.6,0) {\small{$l-1$}};
		\node[below] at (4.2,0) {\small{$l$}};
		\draw [thick] (2.4,0) arc [radius=0.3, start angle=0, end angle= 180];
		\draw [thick] (2.4,.8) arc [radius=0.3, start angle=0, end angle=-180];
	\end{tikzpicture}
\end{align*}
\caption{Graphical representations of the identity operator $\mathbf{1}$ and
the operator $e_i$.}
\end{figure}
\noindent Multiplying two operators corresponds to gluing the picture of the operator sitting on the left  on top
of the picture of the  operator sitting on the right.  Two additional rules also apply. Firstly, two pictures are the same if they
can continuously be deformed into one another. Secondly, a closed loop corresponds to a factor
of $d_{1/2}$. Actually, the first rule ensures  $e_{i} e_{i\pm1} e_{i} = e_i$,  for all $i$, and
$[ e_{i} , e_{j} ] = 0$, for $|i-j| \geqslant 2$, and the second rule  ensures the relation $e_i^2 = d^{\phantom{2}}_{1/2}\, e_i^{\phantom{2}}$.

We now recall briefly how the Temperley--Lieb algebra can be used to show that the
dense anyon model is integrable. For  more details the reader is referred to \cite{feiguin}. 

First, one constructs  a one-parameter family of matrices $R_i(u)$, called $R$-\emph{matrices}, such that each one of them
satisfies the Yang--Baxter equation:
\begin{equation}
\label{eq:ybe}
R_{i} (u) R_{i+1} (u+v)  R_{i} (v) = R_{i+1} (v) R_{i} (u+v) R_{i+1} (u).
\end{equation}
From the $R$-matrices, one in turn constructs  a new one-parameter family of matrices $T(u)$:
\begin{equation}
\label{eq:transfer-matrix}
T(u) := \prod_{i} R_{i}(u),
\end{equation}
called the \emph{transfer} matrices. It follows from Equation~\eqref{eq:ybe}  that $[T(u),T(v)]=0$, for all values of the parameters $u$ and $v$.  Using the transfer matrix $T(u)$, one defines  a Hamiltonian $H$ through 
\begin{align}
	T (u) = \e^{-u\,H + o(u^2)},
\end{align}
from which one obtains the following explicit form for the Hamiltonian $H$:
\begin{align}
\label{eq:ham-from-R}
H = -\frac{\d \ln T(u)}{\d u} \Bigr|_{u=0} =
-\sum_{i} R^{-1}_i (u=0)\,\frac{\d R_i (u)}{\d u}\Bigr|_{u=0}.
\end{align}
By its definition, the Hamiltonian $H$  commutes with the transfer matrices and, hence,  has a large number of
conserved quantities, implying that the model is integrable.

Turning back to the case of the dense anyon model, consider the one-parameter family of matrices $R_i(u)$ defined by
\begin{equation}
\label{eq:r-mat-full}
R_{i} (u) = \sin\Big(\frac{\uppi}{k+2}-u\Big)\,\mathbf{1} + \sin (u)\,e_i.
\end{equation} 
Here, the index $i$ has a similar meaning as in the Hamiltonian, namely, it indicates where
the corresponding matrix acts, and $k$ is the same as in $su(2)_k$.
Using only the algebraic properties of the operators $e_i$  given in \eqref{eqn:TLalgebra}, one can show that the $R_i(u)$ matrices, given by \eqref{eq:r-mat-full},   
satisfy the Yang--Baxter equation. Therefore, from the recipe outlined above, one gets the following Hamiltonian:
\begin{equation}
H = \frac{2}{\tan(\frac{\uppi}{k+2})} \sum_{i} \big(\frac{1}{2}\,\mathbf{1}-\frac{1}{d_{1/2}}\,e_i\big) =
\frac{2}{\tan(\frac{\uppi}{k+2})} \sum_{i} \big(\frac{1}{2}\,\mathbf{1}- h_{i,J=1}\big). 
\end{equation}
This Hamiltonian, up to a shift and a negative overall scale factor, is simply the Hamiltonian $H = \sum_{i} h_{i,J=1}$, which describes the dense anyon chain. Hence,  as mentioned above,  it should be possible to solve the model.
Although doing this turns out to be complicated, however, the solution was obtained
by Andrews, Baxter, and Forrester, who solved the associated two-dimensional
statistical mechanics model \cite{abf}.

We now turn our attention to the dilute anyon model and start by
describing the algebraic structure that we need to construct  $R$-matrices
that satisfy the Yang--Baxter equation in this case. This structure was introduced by
Warnaar et al \cite{warnaar}. Here we follow their presentation.

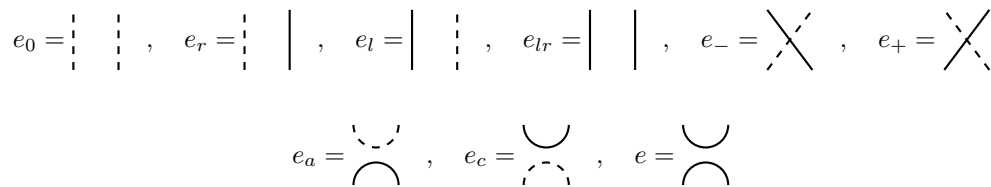
\begin{figure}[H]\centering
\begin{align*}
e_0=\begin{tikzpicture}[baseline=.3cm]
		\draw[thick, dashed] (0,0)--(0,.8);
		\draw[thick, dashed] (.6,0)--(.6,.8);
\end{tikzpicture}\quad,\quad
e_r=\begin{tikzpicture}[baseline=.3cm]
		\draw[thick, dashed] (0,0)--(0,.8);
		\draw[thick] (.6,0)--(.6,.8);
\end{tikzpicture}\quad &,\quad
e_l=\begin{tikzpicture}[baseline=.3cm]
		\draw[thick] (0,0)--(0,.8);
		\draw[thick, dashed] (.6,0)--(.6,.8);
\end{tikzpicture}\quad,\quad
e_{lr}=\begin{tikzpicture}[baseline=.3cm]
		\draw[thick] (0,0)--(0,.8);
		\draw[thick] (.6,0)--(.6,.8);
\end{tikzpicture}\quad,\quad
e_-=\begin{tikzpicture}[baseline=.3cm]
		\draw[thick, dashed] (0,0)--(.6,.8);
		\draw[thick] (.6,0)--(0,.8);
\end{tikzpicture}\quad,\quad
e_+=\begin{tikzpicture}[baseline=.3cm]
		\draw[thick] (0,0)--(.6,.8);
		\draw[thick, dashed] (.6,0)--(0,.8);
\end{tikzpicture}\\\\
	e_a&=\begin{tikzpicture}[baseline=.3cm]
		\draw [thick] (2.4,0) arc [radius=0.3, start angle=0, end angle= 180];
		\draw [thick, dashed] (2.4,.8) arc [radius=0.3, start angle=0, end angle=-180];
	\end{tikzpicture}\quad,\quad
	e_c=\begin{tikzpicture}[baseline=.3cm]
		\draw [thick, dashed] (2.4,0) arc [radius=0.3, start angle=0, end angle= 180];
		\draw [thick] (2.4,.8) arc [radius=0.3, start angle=0, end angle=-180];
	\end{tikzpicture}\quad,\quad
	e=\begin{tikzpicture}[baseline=.3cm]
		\draw [thick] (2.4,0) arc [radius=0.3, start angle=0, end angle= 180];
		\draw [thick] (2.4,.8) arc [radius=0.3, start angle=0, end angle=-180];
	\end{tikzpicture}
\end{align*}
\caption{The nine different types of operators corresponding to the dilute anyon model.}\label{fig:dilute-pictures}
\end{figure}

In the dilute model, sites can be empty.
We represent an empty site by a dashed line. Compared to the dense anyon model, where the only operators considered were the identity operator and $e$, the presence of empty sites for the dilute anyon model provides us with the possibility to consider
additional operators. We introduce these operators by means of their pictorial representation.
For ease of presentation, we give these pictures for a system of two sites only.
For a larger system, one should think that the strands correspond to sites $i$ and $i+1$ and 
 any operator with index $i$ acts as the identity operator on other sites. The operators we need are depicted in Figure~\ref{fig:dilute-pictures}.
\noindent The first four operators $e_{0}$, $e_{r}$, $e_{l}$, and $e_{lr}$ correspond to the identity.
The operators $e_{-}$ and $e_{+}$ move an occupied site one place to the left and
right, respectively. The operators $e_{a}$ and $e_{c}$ correspond to the annihilation and
creation of particles at neighboring sites. Finally, $e$ is the same as before.

Multiplication of the various operators is determined as mentioned earlier, namely, by stacking
pictures on top of one another,  acting from right to left. Hence, the product
$e_\beta e_\alpha$ means that we stack the picture of $e_\beta$ on top of the picture of  $e_\alpha$.
One should note that multiplication of two operators is, by definition, non-zero  only if the dashed and solid lines  corresponding to their pictorial representations match, in the sense that if at some site $i$ from the picture below a dashed (solid) line  is terminated, the line originated from the same site $i$ in the picture above must be a dashed (solid) line as well, for all $i$. For example, $e_{\alpha}e_{+}=0$ for all $\alpha\not=r,-$; in addition, $e_-e_+=e_l$ and $e_re_+=e_+$. We again have the rule that
pictures which can continuously be deformed into each other are equivalent and a 
closed full loop corresponds to a factor $d_{1/2}$. Moreover, in this case, a closed dashed loop can be
removed without any factor or better to say, with a factor $d_{0} = 1$. With the rules above, one can establish the relations
$e_{0} e_{a} = e_{a}$ and $e_{c} e_{0} = e_{c}$, that is, $e_{0}$ acts
as the identity on `matching operators'. The same is true for
$e_{l}$, $e_{r}$, and $e_{lr}$.  Other non-trivial relations with two
operators are
$e_{a} e_{c} = d_{1/2}\,e_{0}$, $e_{a}  e = d_{1/2}\,e_{a}$,
 $e_{c}e_{a} = e$, and $e  e_{c} = d_{1/2}\,e_{c}$.
 
Following a case-by-case-check strategy, we verified that the above-mentioned algebraic relations for $e_{\alpha}$ operators   
are realized by the terms present in the Hamiltonian, provided that one makes the following
identifications:
\begin{align}\label{eqn:identifications}
h_{\mu_{00}} &= \mu_{00}\,e_{0}, &
h_{\mu_{0\frac{1}{2}}} &= \mu_{0\frac{1}{2}}\,e_{r}, &
h_{\mu_{\frac{1}{2}0}} &= \mu_{\frac{1}{2}0}\,e_{l}, &
h_{\mu_{\frac{1}{2}\frac{1}{2}}} &= \mu_{\frac{1}{2}\frac{1}{2}}\,e_{lr},\notag\\
h_{t} &= t\,(e_{-} + e_{+}), &
h'_{\mathit{\Delta}} &= \mathit{\Delta}\, d_{1/2}^{-1/2}\,e_{a}, &
h_{\mathit{\Delta}} &= \mathit{\Delta}\, d_{1/2}^{-1/2 }\,e_{c}, &
h_{J} &= J\, d_{1/2}^{-1}\,e. 
\end{align}
With the algebraic structure in place, we use the results of
reference~\cite{warnaar} to construct  $R$-matrices that satisfy the Yang--Baxter
equation. Consider the following family of matrices: 
\begin{align}\label{eq:r-mat-dilute}
R_{i}(u;\lambda) :=
[\sin(2\lambda)\cos(3\lambda) + \sin(u)\cos(u+3\lambda)]\,e_{i,0} + 
\sin(2\lambda)\cos(u+3\lambda)\,( e_{i,l} + e_{i,r})
\notag\\+ \sin(2\lambda)\sin(u)\,( e_{i,a} + e_{i,c}) + 
\sin(u)\cos(u+3\lambda)\,( e_{i,+} + e_{i,-}) 
\notag\\+ \sin(u+2\lambda)\cos(u+3\lambda)\,e_{i,lr}  + 
\sin(u)\cos(u+\lambda)\,e_{i},
\end{align}
where we reinstated the subscript $i$ on the operators, denoting on which site the
operator acts. One
can verify---using only the algebraic properties of the $e_{i,\alpha}$'s---that these $R$-matrices satisfy the Yang--Baxter Equation~\eqref{eq:ybe},  
provided that $-2\,\cos(4\lambda)=d_{1/2}$.
On the other hand, in our Hamiltonian describing the dilute anyon system, from Equation~\eqref{eq:ds}, 
we get $d_{1/2} = 2\,\cos [\uppi/(k+2)]$. This puts a constraint on
$\lambda$, namely, $\lambda = \pm \uppi\,(k+2 \pm1)/[4(k+2)]$. 

Since the terms in the Hamiltonian satisfy the algebraic relations of the
operators $e_{\alpha}$, one can use the same procedure as outlined above for the dense anyon model 
to find a Hamiltonian that commutes with a set of commuting transfer matrices. Using the $R$-matrices introduced by Equation~\eqref{eq:r-mat-dilute}, the coefficients of the
various terms in the Hamiltonian can be obtained from Equation~\eqref{eq:ham-from-R} and identifications~\eqref{eqn:identifications}. 
The result, up to the overall scaling factor $-\sin(2\lambda)\cos(3\lambda)$, is as follows:
\begin{align}
\label{eq:coefs-integrable}
\nonumber
\mu_{00} &= \cos(3\lambda), & 
\mu_{0\frac{1}{2}} = \mu_{\frac{1}{2}0} &= -\sin(2\lambda)\sin(3\lambda), & 
\mu_{\frac{1}{2}\frac{1}{2}} &= \cos(5\lambda), \\ 
t &= \cos(3\lambda), &
J &= d_{1/2}\,\cos(\lambda), & 
\mathit{\Delta} &= d_{1/2}^{1/2}\,\sin(2\lambda),
\end{align}
where $\lambda$ can take the values $\pm \uppi(k+2 \pm1)/[4(k+2)]$. As mentioned earlier, the system is integrable at the points corresponding to these values for $\lambda$.

As relations in \eqref{eq:coefs-integrable} show, the only coefficient that is odd in $\lambda$ is $\mathit{\Delta}$, which corresponds to the term
in the Hamiltonian that creates and annihilates pairs of anyons. Numeric indicates that the
sign of this term does not change the spectrum of the model. This means that, instead of investigating all four values for $\lambda$,   we
can concentrate ourselves on only two of them, namely,
$\lambda_1 := \uppi\,(k+3)/[4\,(k+2)]$ and $\lambda_2 := \uppi\,(k+1)/[4\,(k+2)]$, provided that we
 consider both the Hamiltonian as defined
by the coefficients as given in \eqref{eq:coefs-integrable}, as well as minus
that Hamiltonian, which of course also commutes with the transfer matrix.

Based on the original paper \cite{warnaar}, for the dilute loop model under consideration there, the points corresponding to these values of $\lambda$  are all critical points. There, using the equivalence to the $O(n)$ model \cite{on1,on2}, the authors have quoted  the values of the central charge $c$ corresponding to
these critical points.
Expressed in terms of the parameter $k$ that is used in this paper, these central charges are given, for each value of $\lambda$, in the following tables:
\begin{align*}
\begin{array}{c|c}
	\lambda & c\\\hline\rule{0pt}{3ex}
	 \frac{\uppi\,(k+3)}{4\,(k+2)} & 1- \frac{6}{(k+2)(k+3)}\\\rule{0pt}{3ex}
	 \frac{\uppi\,(k+1)}{4\,(k+2)} & 1- \frac{6}{(k+1)(k+2)}
\end{array}\qquad\qquad \begin{array}{c|c}
	\lambda & c\\\hline\rule{0pt}{3ex}
	 -\frac{\uppi\,(k+3)}{4\,(k+2)} & \frac{1}{2} + 1- \frac{6}{(k+1)(k+2)}\\\rule{0pt}{3ex}
	 -\frac{\uppi\,(k+1)}{4\,(k+2)} & \frac{1}{2} + 1- \frac{6}{(k+2)(k+3)}
\end{array}
\end{align*}
These central charges point in the direction of minimal-model CFTs.  
Even though the central charges are known in the setting of the dilute loop models
studied in \cite{warnaar}, it is still interesting to investigate the situation for the anyon model
we introduced. This is because the central charge, on its own, does not in general fully determine the corresponding CFT.
We mention that Zhou and Batchelor\cite{zhou97} studied these models using the Bethe Ansatz.

\subsection{Identifying the Critical Points}

In this section, we study the dilute anyon Hamiltonian at the integrable points
that were identified in the previous section. Let $H_1$ and $H_2$ denote the dilute anyon Hamiltonian  corresponding to integrable points $\lambda = \lambda_1$ and  $\lambda = \lambda_2$, 
respectively. 
As mentioned earlier, to investigate all four integrable points determined in the previous section, it suffices to consider the spectra of all four Hamiltonians
$\pm H_{1}$ and $\pm H_{2}$. In each case, using exact diagonalization for small system sizes, we obtain the low-lying part of
the spectrum numerically. Since the dimension of the Hilbert space increases rapidly
 with both system size and $k$, we limit ourselves to $k=1,2,3$.

For a one-dimensional critical system that can be described in terms of a conformal field
theory, the energy of the states, as a function of the system size
$l$ to order $1/l$, takes the following form: 
\begin{equation}
\label{eq:cft-spectrum}
E = E_s l - \frac{\uppi\,v c}{6l} + \frac{2\uppi\,v}{l}\,(2 h_i + n) + \cdots\raisepunct{,}
\end{equation}
where the non-universal constants $E_s$ and $v$ are the energy per site and the
velocity. The constant $c$ denotes the central charge of the conformal field theory,  
 $h_i$'s are the scaling dimensions of the fields of the CFT, and  $n$ is a non-negative
integer. For the primary fields $n=0$, while positive $n$'s give the states corresponding to the
descendants at level $n$. We refer to the references \cite{bpz84,byb} for an introduction on
CFT.

In the case of physical (Hermitian) Hamiltonians, the corresponding CFT is unitary. This means that the
scaling dimension $h_0 = 0$ of the identity field is the smallest scaling dimension. 
In addition, the possible values of the central charge and the possible scaling
dimensions are highly constrained. Therefore, the strategy to identify which CFT (if any) describes
the spectrum at a given integrable point is as follows. 

For a given system size, we obtain the
low-lying part of the spectrum. We then shift and rescale the obtained spectrum in order to
eliminate the non-universal constants in energies given by Equation~\eqref{eq:cft-spectrum}. For the ground state of the system, one has $h_i =n=0$. Hence, the
ground-state energy is given by $E_{0} = E_sl - \frac{\uppi\, v c}{6l}$. We shift the levels by
setting the ground-state energy to zero. The energy $E_1$ of the lowest excited level 
corresponds to the field with the lowest non-zero scaling dimension, which we denote
by $h_{\rm low}$. We rescale the spectrum such that $E_{1} = 1$. After this shift and
rescaling, the energies are given by
\begin{equation}
\label{eq:cft-spectrum-adj}
E = \frac{2 h_i + n}{2h_{\rm low}}\raisepunct{,}
\end{equation}
which does not depend on $E_s$ and $v$ and, therefore, it can be compared with the predictions
for various conformal field theories. Once the CFT has been identified, one can
rescale the energy such that $E_{1} = 2 h_{\rm low}$ and the generic energies
take the form $E = 2 h_i + n$. This rescaling will be used in the plots of the
various energy spectra in the next section. One should note that CFT does not predict at which momenta the
primary fields occur, however, each time $n$ increases by one, the momentum of a state
changes by one, in units of $2\,\uppi/l$, as well. In the next section, we specify at which momenta the
various primary fields occur.

In the case of the dense anyon models, the minimal-model CFTs play an important
role. From the central charges that we quoted in the previous section, we expect that
this will also be the case for the dilute anyon models we consider. We therefore
give the central charge and scaling dimensions of the (unitary) minimal-model CFTs.

The unitary minimal models $\mathcal{M}_m$ are labeled by a parameter $m$ that takes integers greater than or equal to three. The central charge of these minimal models are given by
$c = 1- 6/[m(m+1)]$. The primary fields of $\mathcal{M}_m$, denoted by
$\phi_{r,s}$, are labeled
by two integers $r$ and $s$ with $1\leqslant r \leqslant m$ and $1\leqslant s \leqslant m-1$.
The labels $(r,s)$ and $(m+1-r,m-s)$ correspond to the same field, so $\mathcal{M}_{m}$
has $m(m-1)/2$ primary fields. The scaling dimension $h_{r,s}$ of $\phi_{r,s}$ is given by
\begin{align}
h_{r,s} = \frac{[m r -(m+1)s]^2-1}{4m(m+1)}\raisepunct{.}	
\end{align}

\subsubsection{The integrable point $\lambda_1$}

We start by considering the low-lying part of the spectrum of the
Hamiltonian $H_{1}$, for $k=1,2,3$, and relatively small system sizes.
We find that the spectra, for each value of $k$, can be described in terms of the simple
minimal model $\mathcal{M}_{k+2}$ and, consequently, $\lambda_1$ is a critical point. More explicitly, for $k=1$, the system is
described by the Ising CFT, for $k=2$, the system is described by the tri-critical Ising CFT, and so on.
We mention that in \cite{seaton02}, the CFT describing $H_1$ for $k=2$ was already identified as the
tri-critical Ising CFT. Our findings are consistent with this result. 

\begin{figure}[H]
\includegraphics[width=.45\textwidth]{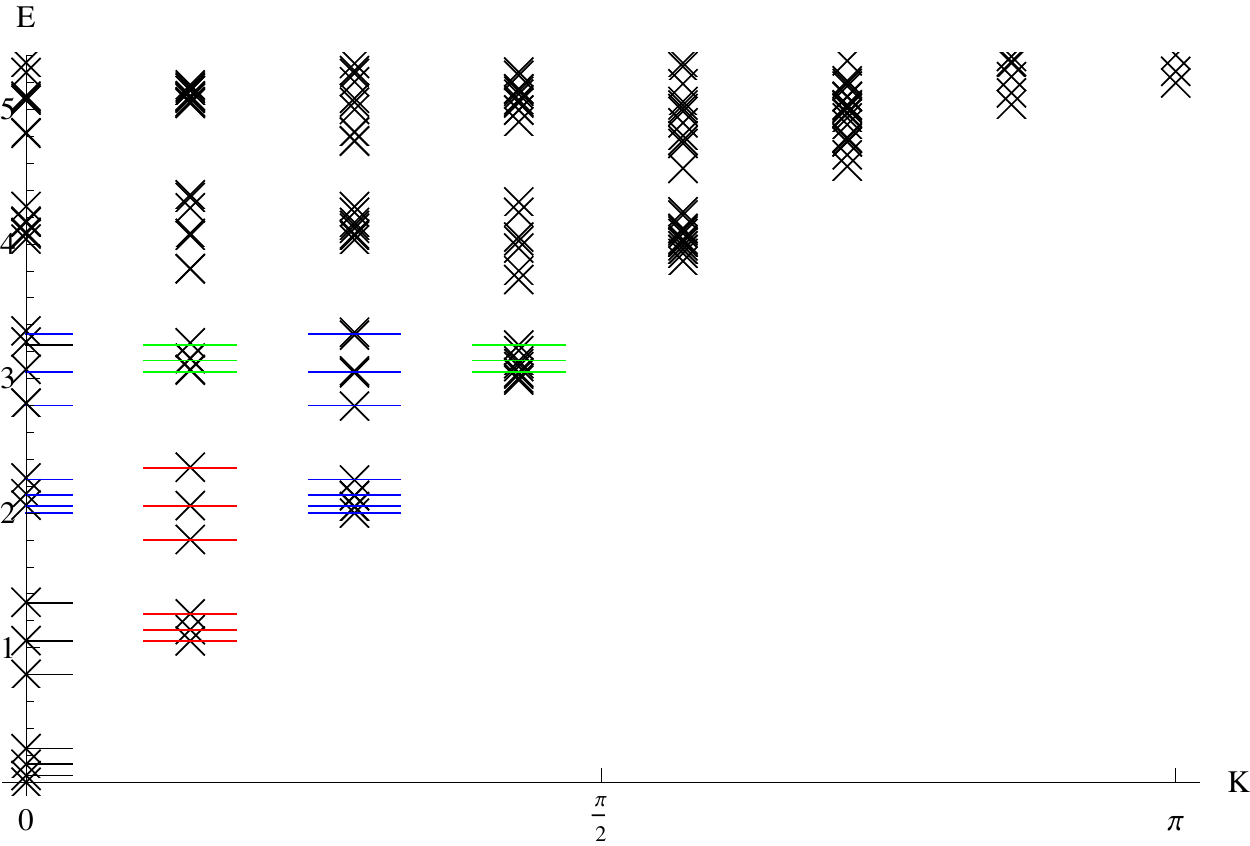}
\hfill
\includegraphics[width=.45\textwidth]{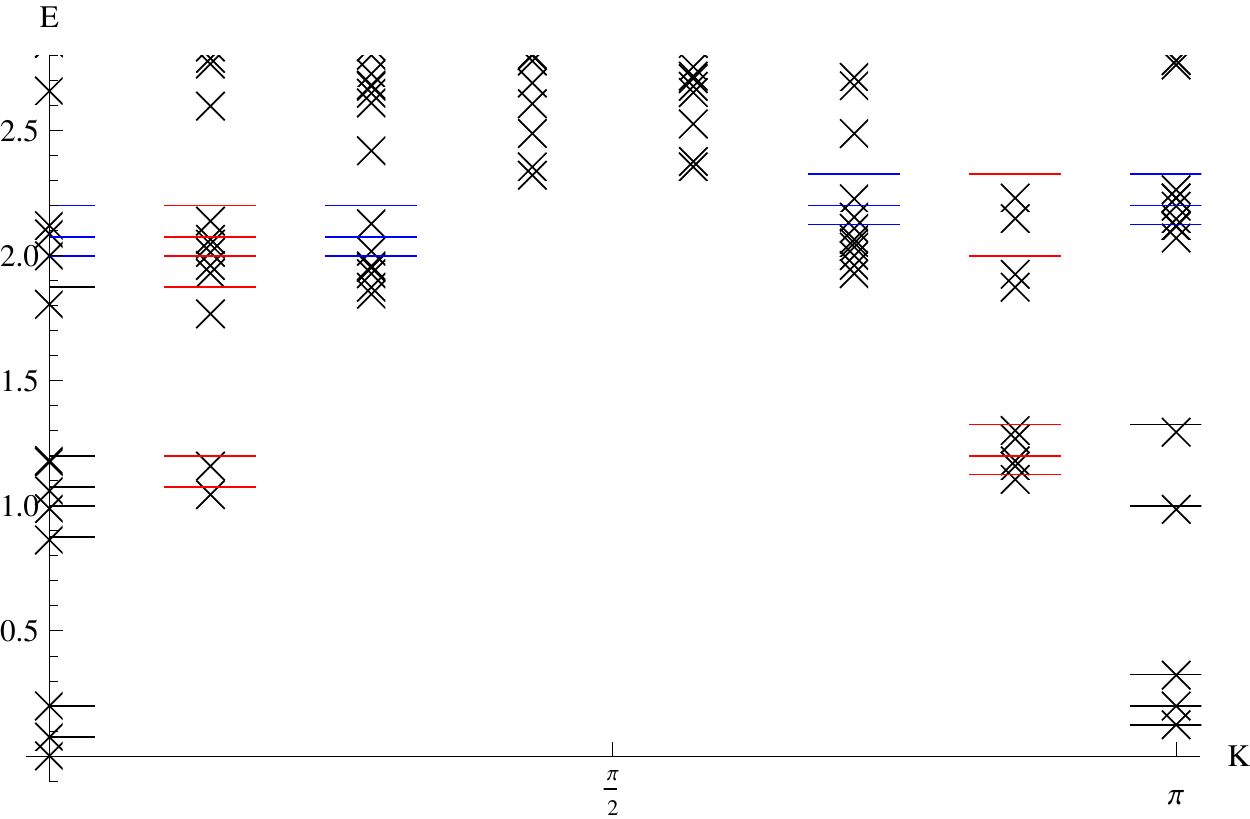}
\caption{
(\emph{Left side})\hspace{.2cm}The spectrum of $H_{1}$, $k=3$ and $l=14$,
and the CFT predictions of the $\mathcal{M}_5$ minimal model.
(\emph{Right side})\hspace{.2cm} The spectrum of $-H_{1}$,  $k=3$ and $l=14$,
 and the CFT predictions of the $({\rm Ising}\times\mathcal{M}_4)$ conformal field theory.}
\label{fig:speck3c1}
\end{figure}

As an illustration, in the left panel of Figure~\ref{fig:speck3c1}, we display the spectrum of
$H_{1}$ for $k=3$ and $l=14$   
as a function of
the momentum $K$. 
\noindent The crosses denote the energy levels, while the
horizontal lines, drawn for ease of comparison, denote the CFT predictions of $\mathcal{M}_5$ for
states with an energy $E < 3.5$. The black lines denote
primary fields, while the red, the blue,  and the green lines denote descendants
at orders $1$, $2$, and $3$, respectively. We observe that, even for this moderate
system size, the numerically obtained spectrum matches the CFT prediction
well. We observe that all the primary fields occur at momentum $K=0$, which   
is also the case for $k=1$ and $k=2$. We expect this to be true for higher $k$'s as well.
We continue with the low-lying part of the spectra of $-H_{1}$, again for
$k=1,2,3$. This time, we could identify the corresponding CFT   
as a product of two minimal models. For $k=1$, the CFT is just $\mathcal{M}_3$, namely, the Ising CFT. We find that, for $k=2$, the CFT is the product of two Ising CFTs and, for
$k=3$, it is the product of the Ising and tri-critical Ising CFTs. Hence, $-\lambda_1$ is also a critical point. In general, the CFT describing this critical point is
$\mathcal{M}_{3} \times \mathcal{M}_{k+1}$, where we identify $\mathcal{M}_2$ with the completely trivial
CFT, that is, the one containing just the vacuum state. 

We display the spectrum of $-H_{1}$, for $k=3$ and $l=14$,  in the
right panel of Figure~\ref{fig:speck3c1}. Despite the rather small system size, we observe a good match with the
$\mathcal{M}_3 \times \mathcal{M}_4$ CFT and
the fact that some of the primary fields occur at momentum $K=\uppi$ that effectively
reduces the system size.

To describe, in general, the momenta of the states corresponding to the primary fields, we
label the fields of the $\mathcal{M}_{3} \times \mathcal{M}_{k+1}$ CFT by $\big(1,(r,s)\big)$, $\big(\sigma,(r,s)\big)$, and
$\big(\psi,(r,s)\big)$, where the first label corresponds to the fields of $\mathcal{M}_{3}$ and the second label 
$(r,s)$ corresponds to the fields of $\mathcal{M}_{k+1}$. So, $1\leqslant r \leqslant k+1$ and $1\leqslant s \leqslant k$.
We observed that the states corresponding to all the primary fields
$\big(1,(r,s)\big)$ and $\big(\psi,(r,s)\big)$ occur at $K=0$, while all the primary fields
$\big(\sigma,(r,s)\big)$ occur at $K=\uppi$.

\subsubsection{The integrable point $\lambda_2$}

For the second integrable point, we could determine whether or not a CFT describes the low-lying
part of the spectrum for $H_{2}$ and we could also identify the corresponding CFT for the cases for which such a CFT exists, but we did not succeed in identifying a
CFT for $-H_{2}$ in general.

\begin{figure}[H]
\begin{center}
\includegraphics[width=.55\textwidth]{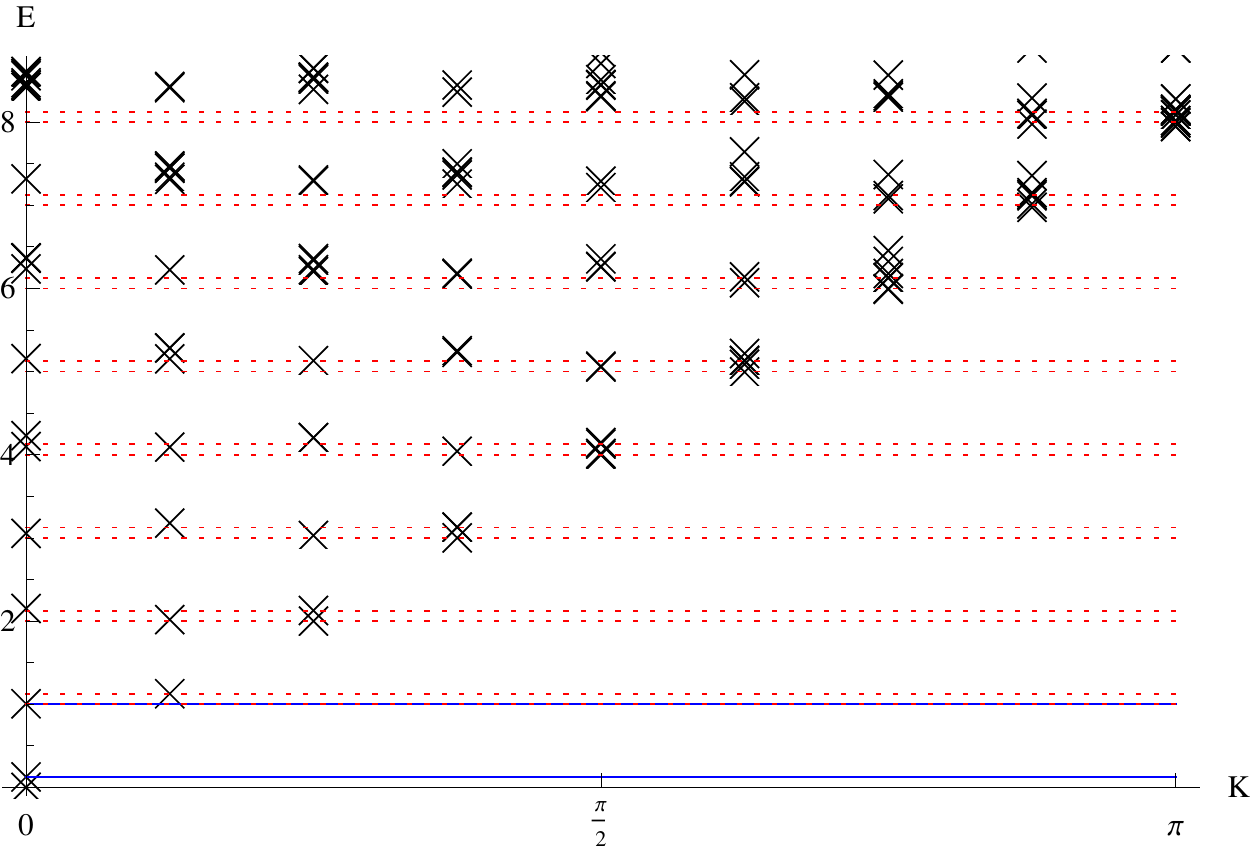}
\end{center}
\caption{
The spectrum of the model Hamiltonian $H_{2}$ for $k=2$ and the CFT predictions of
the ${\rm Ising}$ CFT. The full blue lines indicate the energies corresponding to
the primary fields and the dashed red lines correspond to the descendants.}
\label{fig:speck2c2low}
\end{figure}

By considering the low-lying part of the spectrum of $H_{2}$ for $k=1$, we found
that the spectrum does not allow for a description by a CFT in this case. We 
comment on this below when we discuss the spectrum of $-H_{2}$. 
For
$k=2$, we again find that the spectrum of $H_{2}$ is described by the simple
minimal model $\mathcal{M}_3$, namely, the Ising CFT. 
 In Figure~\ref{fig:speck2c2low},
we give the spectrum for this model for a system with $16$ sites. We observe that, roughly up to energies $E \approx 7$,
this spectrum follows the CFT prediction, which  
is remarkably high for a system of this size. For $k=3$, the spectrum is described well by the minimal model $\mathcal{M}_4$.

Based on these observations, the general picture that results is as follows. For $k \geqslant 2$, the low-lying part of the spectrum of $H_{2}$ is
well described by the minimal model $\mathcal{M}_{k+1}$. Hence, $\lambda_2$ is a critical point, if $k\geqslant2$, and it is not a critical point, if $k=1$. As was the case for the low-lying part of the spectrum of the critical
point of $H_{1}$, we observe that all the states that correspond to the
primary fields have momentum $K=0$. 

Finally, we turn our attention to the case $-H_{2}$. In this case, as indicated above, 
we had problems identifying the correct CFT describing the critical behavior. We focus our
attention on the case $k=1$. Even though the spectrum has features that are reminiscent of
a spectrum described by a CFT, we could not find an obvious match. In principle, there are
many causes that can make identifying the CFT hard. One possible reason is the presence of
large finite-size effect. To gain insight in the situation, we studied the $k=1$ Hamiltonian for
general parameters in more detail. In particular, in Appendix~\ref{app:k1alternative}, we map
this Hamiltonian onto a Hamiltonian for a spin-1/2 chain.

For periodic boundary conditions, the spin-1/2 Hamiltonian one obtains, splits into two sectors.
In both sectors, the number of
down spins is even and the only difference between the two sectors lies in the boundary
conditions. One sector has periodic boundary conditions, while the other sector
has anti-periodic boundary conditions.
Using the parameters corresponding to the integrable point $\lambda_2$, that is, considering
$H_{2}$, we find that the corresponding spin-Hamiltonian takes the form:
\begin{align}
H^{+}_{k=1} &= \sum_{i=1}^{l}
\big(\sigma^x_{i} \sigma^x_{i+1} -
\sigma^y_{i} \sigma^y_{i+1} +
\sigma^z_{i} \sigma^z_{i+1}\big), \\
H^{-}_{k=1} &= \sum_{i=1}^{l}
\big[(-1)^{\delta_{il}}(\sigma^x_{i} \sigma^x_{i+1} -
\sigma^y_{i} \sigma^y_{i+1}) +
\sigma^z_{i} \sigma^z_{i+1}\big].
\end{align}
See Equations~\eqref{eq:k1-spin-periodic} and \eqref{eq:k1-spin-anti-periodic}
in Appendix~\ref{app:k1alternative}. Apart from the difference in the sectors and the minus sign of the $\sigma_i^y \sigma_{i+1}^y$ term, this
is just the ordinary spin-1/2 Heisenberg model. We note that swapping the sign of the
$\sigma_i^y \sigma_{i+1}^y$ term, merely results
in a change of the sign of all the energies, hence, we find that
$H_{2}$ corresponds to $-H_{\rm Heisenberg}$, where
$H_{\rm Heisenberg} = \sum_{i} \bs{\sigma}_i \cdot \bs{\sigma}_{i+1}$.
We are interested in identifying the low-lying part of the spectrum of
$-H_{2}$ that, for $k=1$, corresponds to the low-lying part of the
spectrum of  $H_{\rm Heisenberg}$.

It is known that the finite-size spectra of the Heisenberg model exhibit large finite-size effects,
see for instance \cite{eggert}.
This is the reason that it is hard to determine the correct CFT by means of exact diagonalization
of small systems. Luckily, one can perturb the Heisenberg model in such a way 
that one does not open a gap in the system and, in the meantime,  effectively reduces the finite-size effect. This can be
achieved by adding a Heisenberg term with a next-nearest-neighbor coupling: 
\begin{equation}
H = \sum_{i}( \bs{\sigma}_{i} \cdot \bs{\sigma}_{i+1} +
\lambda\,\bs{\sigma}_{i} \cdot \bs{\sigma}_{i+2}).
\end{equation}
The reader is referred to \cite{eggert} for details. Increasing $\lambda$ from $\lambda = 0$ to $\lambda_c \approx 0.241167$, gradually reduces the
finite-size effects. For $\lambda > \lambda_c$, the system becomes gapped and enters
the Majumdar-Ghosh phase \cite{mga,mgb}.

Based on this, we can reduce the finite-size effects by studying the dilute anyon model that
includes both nearest-neighbor and  next-nearest-neighbor terms. We did this for the
$k=1$ anyon model, which corresponds to the Heisenberg spin-1/2 chain by taking the parameters
for the nearest-neighbor terms to be $\mu_{0} = 0$, $\mu_{1}=\mu_{2}= -1$,
$t=1$, $\mathit{\Delta}=0$, and $J=1$.  For the notations used for these parameters, see the last two lines of the forth paragraph of Appendix~\ref{app:k1alternative}. The low-lying part of the spectrum of this model corresponds
to the low-lying part of the spectrum of $-H_{2}$. 
The parameters of the next-nearest-neighbor terms are obtained from the
nearest-neighbor ones by multiplying them with $\lambda_c$.

\begin{figure}[H]
\begin{center}
\includegraphics[width=.55\textwidth]{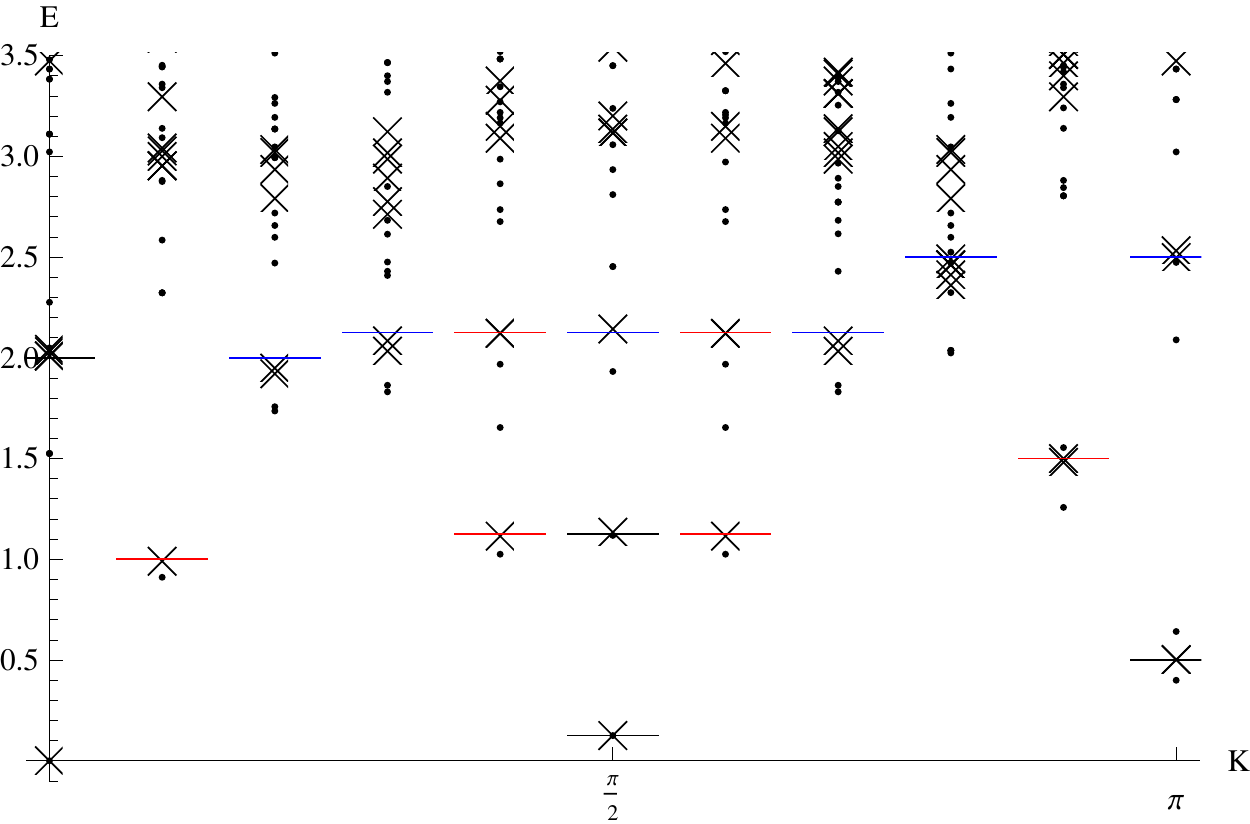}
\end{center}
\caption{
The spectrum of the model Hamiltonian $-H_{2}$  for $k=1$ with next-nearest-neighbor terms
(crosses) as well as with only nearest-neighbor terms (dots). 
The CFT predictions for the $u(1)_8$ CFT are indicated by the black 
lines (primary fields), the red  lines (first descendants), and the blue lines (second descendants).}
\label{fig:speck2c2high}
\end{figure}

In Figure~\ref{fig:speck2c2high}, we show the spectrum of this model for a system of size $l=20$.
The energy levels are denoted by the crosses. For comparison,
we also include the states of the model without the next-nearest-neighbor terms, which
are denoted by small dots. One observes that
the effect of adding the next-nearest-neighbor terms does change the spectrum significantly.

From the spectrum including the next-nearest-neighbor terms, we can identify the
CFT that describes the critical behavior of the $-H_{2}$ model with $k=1$.
The CFT is a compactified boson with $c=1$ and has eight primary fields 
whose scaling dimensions are $h_p =p^2/16$, where $p =0,\pm 1,\pm 2,\pm 3,4$.
The momenta of the states corresponding to the primary fields are given by $p\uppi/2\,(\bmod\, 2\uppi)$.
In Figure~\ref{fig:speck2c2high}, for states with
$E \leqslant 2.5$, we also indicate the CFT primary fields (black lines), first
descendants (red lines), and second descendants (blue lines). We find that the numerically obtained energies match this CFT spectrum well.

To provide further evidence that we correctly identified the CFT, we give the prediction of the
multiplicities of the states for this CFT and the closely related one, namely, the four-state Potts CFT.
In the case of
particular twisted boundary conditions, this CFT describes the critical behavior of the spin-1/2 Heisenberg model~\cite{abb88}.  
The four-state Potts CFT
also has $c=1$ and can be viewed as the `orbifold' of the compactified boson CFT with eight primary
fields. The multiplicities can be obtained from the characters of these CFTs~\cite{dvv88}. These multiplicities together with the numerically obtained multiplicities for $-H_2$ with $k=1$ are
given in Table~\ref{tab:multiplicities}. We find perfect agreement with the $u(1)_8$ CFT, while the 4-state Potts CFT , even for the lowest-lying states, 
does not give the right multiplicities. 
\begin{table}[ht]
\begin{center}
\begin{tabular}{r|c|c|c|c|c|c|c|c|c}
$E$ & $0$ & $1/8$ & $1/2$ & $1$ & $9/8$ & $3/2$ & $2$ & $17/8$ & $5/2$\\
\hline
$u(1)_8$ & $1$ & $2$ & $2$ & $2$ & $6$ & $4$ & $9$ & $18$ & $14$\\
\hline 
4-state Potts & $1$ & $3$ & $1$ & $0$ & $9$ & $2$ & $5$ & $27$ & $7$\\
\hline 
$-H_{2}$ for $k=1$ & $1$ & $2$ & $2$ & $2$ & $6$ & $4$ & $9$ & $18$ & $14$\\
\end{tabular}
\end{center}
\caption{The multiplicities predicted by the $u(1)_8$ and 4-state Potts CFTs, as well as the
numerically obtained multiplicities for $-H_2$ for $k=1$.}
\label{tab:multiplicities}
\end{table}

Given that it was hard to determine the CFT describing the critical behavior of the
model Hamiltonian  $-H_{2}$ for $k=1$, it is not surprising that it is hard to figure out the CFT
for higher values of $k$. We leave this for future work.

We close the discussion on the integrable point $\lambda_2$ by commenting on the
observation that the spectrum of the model Hamiltonian $H_{2}$ for $k=1$ is not described by a CFT.
By using the mapping to the spin model, we found that this spectrum is described by the
ferro-magnetic spin-1/2 Heisenberg model. 
It is known that this model is gapless, but the excitations, so-called spin waves,
 have a quadratic dispersion instead of the linear dispersion that is predicted by CFT.

To conclude this section, we mention that we were able to identify the CFT that describes
the spectra of the Hamiltonians $H_{1}$ and $-H_{1}$ for $k\geqslant 1$, and the Hamiltonian $H_{2}$ for $k\geqslant 2$.
Not surprisingly, in each case, the central charge of the CFT matches the central charges
obtained in \cite{warnaar} for the dilute loop models.
For the Hamiltonian $-H_{2}$, the analysis is hampered by large finite-size effects.
Nevertheless, we determined the CFT describing this model, for $k=1$, as compactified
boson CFT with central charge equal to one. Again, the central charge matches the value obtained for the dilute loop model.
However, it was not clear that which  CFT with central charge equal to one would be the right description. Other possibilities
would have been the product of two Ising CFTs or the 4-state Potts CFT. In light of this, it is interesting to investigate
this model for higher values of $k$ in more detail.


\section{Conclusion}
\label{sec:conclusions}
In this paper we introduced a one-dimensional anyon model with the feature
that the number of anyons, the spin-1/2 anyon of $su(2)_k$, is allowed to fluctuate.
This is achieved by adding a
pairing term that creates and annihilates pairs of anyons on neighboring sites.
This term is analogous to the pairing term that creates Cooper pairs in mean-field
description of superconductors, such as the Kitaev chain model of a one-dimensions
$p$-wave superconductor.

We studied the model at five special points. At the first point, the model is a sum
of projectors. For $k$ odd, there are exact zero-energy ground states which indicates
that the model is gapped. For even $k$, there are no such exact zero-energy ground states 
and it would be interesting to study the model in more detail to determine if the model
is also gapped at this point.

At the other four points, the system is integrable for all values of $k$.
At these points, corresponding to four different choices of the parameters,
the model is gapless and we were able to determine the CFT description
in three out of the four cases. In the remaining case, large finite-size effects
were the cause that exact diagonalization of small system sizes does not
yield enough information to determine the CFT. For $k=1$, however, the
Hamiltonian maps to a particular version of the spin-1/2 Heisenberg for which
one can reduce the finite-size effects by adding next-nearest-neighbor terms.
It would be interesting to see if one can reduce the finite-size effects in a similar
way for higher values of $k$.

As is already noted in \cite{feiguin}, the anyon models are closely related to
two-dimensional statistical-mechanics-models, namely the
so-called `restricted solid on solid models' (RSOS), which were introduced by
Andrews, Baxter and Forrester\cite{abf}. Using this connection, it is possible to
obtain information about the critical point of the anyon models.
RSOS models have attracted much attention over the years and various
generalizations have been considered, see for instance\cite{date86,pasquier87,warnaar}.
Much more recently\cite{kakashvili}, a generalized RSOS model was constructed
to solve an integrable point of the anyon models considered in \cite{anyon-long-range}.
It would be interesting to see if the techniques introduced by Andrews, Baxter and Forrester
can by employed to shed light on the critical point for which we could not identify the
CFT for arbitrary $k$.

Finally, we mention that we did not embark on a more detailed numerical study of the
model, but we hope that the behavior at the special points mentioned above can be a
guide for such a study.

\nopagebreak[4]

\section*{Acknowledgements}
E.A. would like to thank M. Soni and D. Poilblanc for discussions.  This work was sponsored,
in part, by the Swedish research council.


\appendix
\chapter{The $F$-symbols}
\label{app:f-symbols}

The form we use for $F$-symbols in this article deviates slightly from the `standard form'. 
In this appendix, we explicitly give the $F$-symbols for the $su(2)_k$ fusion rules that we use in this
paper. In particular,
it is convenient for our purposes that the $F$-symbols we use in the Hamiltonian are all positive.

We denote the standard form of the $F$-symbols by $\widetilde{F}$,
which is derived in~\cite{kr88,pasquier}, and we give it here for completeness:
\begin{align}\label{eqn:fsymbol}
	\widetilde{F}^{abc}_{d;\,ef}&=(-1)^{a+b+c+d}\mathit{\mathit{\Delta}}(a,b,e)\mathit{\mathit{\Delta}}(c,d,e)\mathit{\mathit{\Delta}}(b,c,f)\mathit{\mathit{\Delta}}(a,d,f)\sqrt{\lfloor 2e+1\rfloor_q}\sqrt{\lfloor 2f+1\rfloor_q}\notag\\&
	\hspace{.6cm}\times\sum_{n=m}^{M}(-1)^n\bigg(\frac{\lfloor n+1\rfloor_q!}{\lfloor a+b+c+d-n\rfloor_q!\lfloor a+c+e+f-n\rfloor_q!\lfloor b+d+e+f-n\rfloor_q!}\notag
	\\&\hspace{3.1cm}\times\frac{1}{\lfloor n-a-b-e\rfloor_q!\lfloor n-c-d-e\rfloor_q!\lfloor n-b-c-f\rfloor_q!\lfloor n-a-d-f\rfloor_q!}\bigg)\raisepunct{.}
\end{align}
Here for any real number $r$, the so-called $q$-number $\lfloor r\rfloor_q$, is defined by
\begin{align}
	\lfloor r\rfloor_q=\frac{q^{r/2}-q^{-r/2}}{q^{1/2}-q^{-1/2}}\raisepunct{,}\quad q:=\exp\Big(\frac{2\uppi\,\i}{k+2}\Big)\raisepunct{,}
\end{align}
and
for a non-negative integer $n$, the $q$-factorial $\lfloor n\rfloor_q!$ is defined by
\begin{align}
	\lfloor n\rfloor_q!=\lfloor n\rfloor_q\lfloor n-1\rfloor_q\cdots\lfloor 1\rfloor_q,\quad\lfloor 0\rfloor_q!:=1.
\end{align}
Moreover, for labels  $a$, $b$, and $c$ from $\{0,1/2,\ldots,k/2\}$, with $a\leqslant b+c$, $b\leqslant c+a$, $a\leqslant b+c$, and $a+b+c=0\;(\hspace{-.2cm}\mod 1)$,
\begin{align}
	\mathit{\Delta}(a,b,c):=\sqrt{\frac{\lfloor a+b-c\rfloor_q!\lfloor a-b+c\rfloor_q!\lfloor -a+b+c\rfloor_q!}{\lfloor a+b+c+1\rfloor_q!}}\raisepunct{.}
\end{align}
Finally, summation-limits $m$ and $M$ are defined by
\begin{align}
	m&=\max\{a+b+e,c+d+e,b+c+f,a+d+f\}\raisepunct{,}\\
	M&=\min\{a+b+c+d,a+c+e+f,b+d+e+f\}\raisepunct{.}
\end{align}
These values for summation-limits guarantees that the arguments of $q$-factorials appeared in Equation~\eqref{eqn:fsymbol} to be non-negative integers. The quantum dimensions $d_j$ relate themselves to the notion of $q$-numbers through the relation $d_j=\lfloor 2j+1\rfloor_q$. For the solutions given by Equation~\eqref{eqn:fsymbol}, the $\widetilde{F}$-matrices are their own inverses.

The $F$-symbols that appear in the Hamiltonian of our model take the form
$\widetilde{F}^{\,x\frac12\frac12}_{x;\,x\pm\frac12,0}$. From Equation~\eqref{eqn:fsymbol}, one finds: 
\begin{equation}
\widetilde{F}^{\,x\frac12\frac12}_{\,x} =
\frac{1}{\sqrt{d_{1/2}d_{x}}}\begin{bmatrix}
-\sqrt{d_{x-1/2}} & \sqrt{d_{x+1/2}}\;\; \\\\
\sqrt{d_{x+1/2}} & \sqrt{d_{x-1/2}}\;\;
\end{bmatrix},\qquad(0<x<k/2),
\end{equation}
where the rows correspond to $e = x-1/2,x+1/2$ and the columns to $f=0,1$, respectively.

In general, the $F$-symbols have the following gauge freedom. If a set of
$F$-symbols $\widetilde{F}^{abc}_{d;\,ef}$ is a solution to the Pentagon equations, then the set of $F$-symbols $F^{abc}_{d;\,ef}$  defined by
\begin{align}
F^{abc}_{d;\,ef} = \frac{u^{bc}_{f}u^{af}_d}{u^{ab}_{e}u^{ec}_{d}}\,\widetilde{F}^{abc}_{d;\,ef},
\end{align}
where $u^{ab}_{c}$ are arbitrary constants and are called \emph{gauge factors}, is an \emph{equivalent} solution---equivalent in the sense that although the new set of $F$-symbols change the
explicit form of the Hamiltonian in general, but the new Hamiltonian  has the same spectrum as the previous one.

In this paper, all gauge factors have been assigned either $-1$ or $+1$. We have assigned $-1$ to gauge factors of the forms $u^{a0}_{a}$ and $u^{a+\frac12,\frac12}_{a}$, if $a$ is a half-integer,  and also to gauge factors of the form $u^{a1}_{a}$, if $a$ is an integer.  In all other cases, the gauge factors have taken to be  $+1$.
One can check that in this gauge, all the $F$-symbols that appear in the Hamiltonian are
positive. More explicitly, we have:
\begin{equation}
F^{\,x\frac12\frac12}_{x} =
\frac{1}{\sqrt{d_{1/2}d_{x}}}\begin{bmatrix}
\sqrt{d_{x-1/2}} & \sqrt{d_{x+1/2}}\;\; \\\\
\sqrt{d_{x+1/2}} & -\sqrt{d_{x-1/2}}\;\;
\end{bmatrix}.
\end{equation}
Again the rows correspond to $e = x-1/2,x+1/2$ and the columns to $f=0,1$, respectively.

\chapter{Alternative Formulation of the $k=1$ Chain}
\label{app:k1alternative}

In this appendix, we write the anyon chain model for $k=1$ in terms of a
spin-1/2 model. In this way, we hope to get insight in the critical behavior at the
second critical point that we identified. The spin-1/2 model obtained in this way is a version 
of the XYZ model in a magnetic field, whose phase diagram has been investigated,
see for instance \cite{peschel81,kurmann82}.

First, we consider the chain with open boundary conditions and start by comparing the
Hilbert spaces of the anyon chain and a spin-1/2 chain, both with $l$ sites. We know that the Hilbert space of a spin-$1/2$ chain with $l$ sites has dimension $2^l$.
For the anyon chain with $l$ sites, each of the sites can be occupied with an anyon or
be empty, namely, $y_i = 0, 1/2$ for $i=1,\ldots,l$.
In addition, we have to take the labels of the fusion chain, $x_i$'s,
into account. For $k=1$, the fusion rules are Abelian and, thus, all the labels $x_i$, with
$i = 1, 2,\ldots,l$, are specified, once $x_0$ is specified. Therefore, if we allow for all possibilities,
the Hilbert space of the anyon chain has dimension $2^{l+1}$. We note,  however,  that the anyon
chains with $x_0=0$ and $x_0 = 1/2$ are completely equivalent. Therefore, we simply consider the
chain with $x_0 = 0$ and rewrite this chain in terms of a spin-1/2 chain.

To map the anyon chain to a spin chain, we first need to find a correspondence between the 
anyon and spin degrees of freedom. For the open chain, it turns out that the simplest possible
correspondence, in which an empty site $y_i = 0$ corresponds to a spin-up
$| \uparrow_i \rangle$ and an occupied site $y_i = 1/2$ corresponds to a spin-down
$| \downarrow_i \rangle$, works.

With these conventions in place, we can start to write the Hamiltonian for the anyon
chain in terms of the Hamiltonian for the spin chain. 
 We deal with the terms that act diagonally first. For $k=1$, this includes the
`chemical-potential' terms as well as the `interaction term' $h_J$. Note that, for $k=1$, the interaction term $h_J$
acts diagonally. In fact, it acts in exactly the same way as the term $h_{\mu_{\frac{1}{2}\frac{1}{2}}}$ does, namely, it
 assigns energy only if two neighboring sites are occupied. Hence, we can easily take
the $h_J$ term into account by combining it with the term $h_{\mu_{\frac{1}{2}\frac{1}{2}}}$. For simplicity, we consider the case $\mu_{1}:=\mu_{0\frac{1}{2}} = \mu_{\frac{1}{2}0}$,
where the subscript $1$ denotes that only one of the two sites is occupied. In this notation, we
also have $\mu_{0}:=\mu_{00}$ and $\mu_{2}:=\mu_{\frac{1}{2}\frac{1}{2}}$.

The diagonal term in the anyon chain acting on $(y_i,y_{i+1})$ assigns
an energy $\mu_{0}$, if $(y_i,y_{i+1}) = (0,0)$,
assigns an energy $\mu_{1}$, if $(y_i,y_{i+1}) = (0,1/2)$ or $(1/2,0)$, and assigns
an energy $\mu_{2} + J$, if $(y_i,y_{i+1}) = (1/2,1/2)$.
The most general spin-term acting on the neighboring sites $i$ and $i+1$ takes the form
$\alpha\,\sigma_{i}^z \sigma_{i+1}^z + \beta\,\sigma_{i}^z +
\beta'\,\sigma_{i+1}^z + \gamma\,\mathbf{1}$. Matching the coefficients gives
$\alpha = (\mu_{0}-2\mu_{1}+\mu_{2}+J)/4$,
$\beta = \beta' = (\mu_{0}-\mu_{2}-J)/4$, and
$\gamma = (\mu_{0}+2\mu_{1}+\mu_{2}+J)/4$.
At the cost of introducing two terms that act on the
first and last sites, we can separate the $\sigma_{i}^z \sigma_{i+1}^z$ term, which acts on
neighboring sites, from the other ones. This gives the diagonal part $H_{d}$ of the Hamiltonian:
\begin{align}
H_{d} &=\smash[b]{ \sum_{i=1}^{l}}
\frac{\mu_{0} -\mu_{2} - J}{2}\,\sigma_{i}^z 
\nonumber
+ \smash[b]{\sum_{i=1}^{l - 1}}\frac{\mu_{0} - 2\mu_{1} + \mu_{2} + J}{4}\,\sigma_{i}^z \sigma_{i+1}^z 
\nonumber
\\&\hspace{4cm}-  \frac{\mu_{0} -\mu_{2} - J}{4}\,(\sigma_{1}^z + \sigma_{L}^z)
- \frac{l-1}{4} (\mu_{0} + 2\mu_{1} + \mu_{2} + J)\, \mathbf{1}.
\end{align}

Finally, we need to consider the off-diagonal terms in the Hamiltonian, namely,  the hopping
and `superconducting' terms. The hopping term, which hops an anyon from site $i$ to
$i+1$ or vice-versa, takes the form 
$\sigma^{-}_{i} \sigma^{+}_{i+1}+\sigma^{+}_{i} \sigma^{-}_{i+1}$, in terms of the spin raising and lowering operators, or, equivalently, takes the form 
$1/2(\sigma^{x}_{i} \sigma^{x}_{i+1}+\sigma^{y}_{i} \sigma^{y}_{i+1})$, in terms of spin operators. Similarly, the term that
creates or annihilates a pair of anyons on two neighboring sites, takes the form
$\sigma^{+}_{i} \sigma^{+}_{i+1} + \sigma^{-}_{i} \sigma^{-}_{i+1}$ or
$1/2(\sigma^{x}_{i} \sigma^{x}_{i+1} - \sigma^{y}_{i} \sigma^{y}_{i+1})$.
Thus, the final form of the spin Hamiltonian that is equivalent to the open $k=1$ anyon chain has the following form:
\begin{align}
H_{k=1,{\rm spin}} =&
\smash[b]{\sum_{i=1}^{l}}
\frac{\mu_{0} -\mu_{2} - J}{2}\, \sigma_{i}^z \notag\\&
\hspace{1cm}+  \smash[b]{\sum_{i=1}^{l - 1}}\Big( \frac{\mu_{0} - 2\mu_{1} + \mu_{2} + J}{4}\, \sigma_{i}^z \sigma_{i+1}^z
+ \frac{t+\mathit{\Delta}}{2}\, \sigma_{i}^x \sigma_{i+1}^x + \frac{t-\mathit{\Delta}}{2}\, \sigma_{i}^y \sigma_{i+1}^y\Big) \notag\\&
\hspace{3.7cm}-\frac{\mu_{0} -\mu_{2} - J}{4}\, (\sigma_{1}^z + \sigma_{L}^z)
+ \frac{l-1}{4}(\mu_{0} + 2\mu_{1} + \mu_{2} + J)\,\mathbf{1}.
\end{align}

We now turn our attention to the anyon chain with periodic boundary conditions.
The main difference with the open case is that the relation between the anyon and
spin Hilbert spaces is a bit more complicated. The Hilbert space of the periodic spin system
is identical to the Hilbert space of the open spin system and has dimension $2^l$. In the
case of the anyon chain with periodic boundary conditions, there is a constraint on the
number of anyons in the system. Namely, because of the fusion rules, the number of anyons
in the system, that is, the number of $y_{i}$ labels that take the value $1/2$, has to be even. This gives
us $2^{l-1}$ possible assignments for the $y_{i}$ labels. However, for each assignment to the $y_i$ labels, there are
now two distinct assignments for the $x_{i}$ labels consistent with fusion rules. The $x_{i}$ labels take the values $0$ and $1/2$ and the
relation between the consistent assignments for the $x_{i}$ labels sends $0$ to  $1/2$ and vice-versa.
 Thus, the total dimension of the Hilbert space is again given by $2^{l}$, just
as for the spin system.

To relate the degrees of freedom between the anyon and spin system, we run into a problem.
In the open case, we identified an empty anyon site with a spin-up and a filled anyon site
with a spin-down. In the periodic case, we can only have an even number of anyons, which
corresponds to an even number of spin-downs in the spin system. For a spin chain of length $l$,
there are only $2^{l-1}$ such states which means that we have twice as many anyon states
 compared to the number of spin states. To deal with this problem, we first take a closer
look at the anyon system.

As we stated above, in the anyon chain with periodic boundary conditions, the number of
anyons has to be even,  giving $2^{l-1}$ assignments for the $y_{i}$ labels. We can divide the
corresponding assignments for the $x_{i}$ labels in two classes, namely, those with $x_{0} = 0$, and
those with $x_{0} = 1/2$. One should note that the choice of $x_{0}$, as the label to make the distinction,  is
arbitrary and one could have well chosen any other label. We denote these two sets of states by
$|\psi_{j,x_{0} = 0} \rangle$ and $|\psi_{j,x_{0}=1/2} \rangle$, where $j$ labels the
$2^{l-1}$ states in each
set. For any given $j$, the state $|\psi_{j,x_{0}=1/2} \rangle$ can be obtained from
$|\psi_{j,x_{0}=0}\rangle$ by sending $x_{i}$ to $1/2 - x_{i}$, that is, by the exchange
$0 \leftrightarrow 1/2$ for each $x_{i}$.

We note that the Hamiltonian can change the value of $x_{0}$ by hopping an
anyon `over the boundary' or by creating or annihilating a pair of anyons on sites
$1$ and $l$. Thus, the Hamiltonian mixes the two sets of states
$|\psi_{j,x_{0}=0} \rangle$ and $|\psi_{j,x_{0}=1/2} \rangle$. It is possible, however, to go
to a basis in which the Hamiltonian becomes block-diagonal, with two blocks of  size
$2^{l-1}$ each. As we now discuss, the basis in which this happens is
\begin{align}
| \psi^{+}_j \rangle := \frac{1}{\sqrt{2}} \bigl(\,
|\psi_{j,x_{0}=0} \rangle + |\psi_{j,x_{0}=1/2} \rangle
\bigr),\quad
| \psi^{-}_j \rangle := \frac{1}{\sqrt{2}} \bigl(\,
|\psi_{j,x_{0}=0} \rangle - |\psi_{j,x_{0}=1/2} \rangle
\bigr).
\end{align}
To see that in this basis the Hamiltonian splits up in two blocks, we note the following.
Only the term in the Hamiltonian that hops an anyon `over the boundary'---denoted
here by $h_{t,l}$---and the terms that create or
annihilate a pair of anyons `over the boundary'---denoted here by $h_{\mathit{\Delta},l}$ and $h'_{\mathit{\Delta},l}$, respectively---can change the label $x_0$. In addition,
all the other terms act in the same way on
$|\psi_{j,x_{0}=0} \rangle$ and $|\psi_{j,x_{0}=1/2} \rangle$.
The action of $h_{t,l}$ on 
$|\psi_{j,x_{0}=0} \rangle$ gives one state with $x_{0}=1/2$, say
$t\,|\psi_{j',x_{0}=1/2} \rangle$. Then, we have that
$h_{t,l} |\psi_{j,x_{0}=1/2}\rangle  = t\,|\psi_{j',x_{0}=0}\rangle$. Thus, we find that
$h_{t,l} | \psi^{+}_j \rangle  = t\,| \psi^{+}_j \rangle$ and
$h_{t,l} | \psi^{-}_j \rangle  =  -t\,| \psi^{+}_j \rangle$. Indeed, the term
$h_{t,l}$ does not mix the sectors $| \psi^{+}_j \rangle$ and
$| \psi^{-}_j \rangle$. Furthermore,  
 $h_{t,l}$ acts on $| \psi^{-}_j \rangle$ with an additional minus sign. Exactly the same reasoning
applies to $h_{\mathit{\Delta},l}$ and $h'_{\mathit{\Delta},l}$ terms.

We can use the decomposition of the anyon Hamiltonian into the two blocks mentioned above to find the
corresponding spin Hamiltonian. In both blocks, the number of anyons is even and the only
difference in the form of the Hamiltonian is the additional sign in the hopping, creation, and
annihilation terms going across the boundary. This means that we can map the anyon
Hamiltonian to a spin system in the following way. We consider a spin system with an even
number of spin-downs, that is, a spin system with a Hilbert space of dimension $2^{l-1}$. The spectrum that corresponds
to the spectrum of the anyon Hamiltonian acting on the space $| \psi^{+}_j \rangle$ can be found
in the same way as we did in the open case, resulting in the following  Hamiltonian:
\begin{align}
H_{k=1}^{+} =& 
\smash[b]{\sum_{i=1}^{l}} \Big[(\mu_{0}-\mu_{1})\,\sigma_i^z + \mu_{1}\,\mathbf{1}
+ \frac{t+\mathit{\Delta}}{2}\, \sigma_{i}^x \sigma_{i+1}^x + \frac{t-\mathit{\Delta}}{2}\,\sigma_{i}^y \sigma_{i+1}^y \notag\\
& \hspace{3.5cm}+ \frac{\mu_{2} + J - 2\mu_{1} + \mu_{0}}{4}\, 
(\sigma_{i}^z \sigma_{i+1}^z - \sigma_{i}^z - \sigma_{i+1}^z + \mathbf{1})\Big],
\end{align}
which by collecting the $\sigma_i^z$ terms,  becomes:
\begin{align}
\label{eq:k1-spin-periodic}
H_{k=1}^{+} =& 
\smash[b]{\sum_{i=1}^{l}} \Big(\frac{\mu_{0}-\mu_{2}-J}{2}\,\sigma_i^z 
+ \frac{t+\mathit{\Delta}}{2}\,\sigma_{i}^x \sigma_{i+1}^x + \frac{t-\mathit{\Delta}}{2}\,\sigma_{i}^y \sigma_{i+1}^y \notag\\
& \hspace{3cm}+ \frac{\mu_{2} + J - 2\mu_{1} + \mu_{0}}{4}\,\sigma_{i}^z \sigma_{i+1}^z
+ \frac{\mu_{0}+2\mu_{1}+\mu_{2} + J}{4}\,\mathbf{1}\Big).
\end{align}
The anyon spectrum in the space $| \psi^{-}_j \rangle$ corresponds to the spectrum of the
following spin Hamiltonian:  
\begin{align}
\label{eq:k1-spin-anti-periodic}
H_{k=1}^{-} =& 
\smash[b]{\sum_{i=1}^{l}} \Big(\frac{\mu_{0}-\mu_{2}-J}{2}\,\sigma_i^z 
+ (-1)^{\delta_{il}}\,\frac{t+\mathit{\Delta}}{2}\,\sigma_{i}^x \sigma_{i+1}^x +
(-1)^{\delta_{il}}\,\frac{t-\mathit{\Delta}}{2}\, \sigma_{i}^y \sigma_{i+1}^y\notag \\
& \hspace{4cm}+ \frac{\mu_{2} + J - 2\mu_{1} + \mu_{0}}{4}\,\sigma_{i}^z \sigma_{i+1}^z
+ \frac{\mu_{0}+2\mu_{1}+\mu_{2} + J}{4}\,\mathbf{1}\Big),
\end{align}
which also acts on the space in which all states have an even number of
spin-downs and whose only difference with the Hamiltonian $H_{k=1}^{+}$ is the change in boundary
conditions. We note that there is no additional sign for the term $\sigma_{l}^z \sigma_{1}^z$,
because it acts diagonally.

In conclusion, we found that the $k=1$ anyon Hamiltonian can be written in terms of the
$XYZ$ spin-1/2 Hamiltonian with a magnetic field in the $z$ direction. In the case of an open
anyon chain, there are boundary terms, while in the periodic case, the spectrum corresponds
to two versions of a spin model. Both versions have an even number of spin-downs but have
different boundary conditions. For concreteness, we give the explicit form of the spin-1/2 Hamiltonian in the sector with
periodic boundary conditions. For the two critical points we have:
\begin{align}
H_{k=1,{\rm c1}} &=\sum_{i}
\Big(\frac{\sqrt{3}-2}{4}\,\sigma^{x}_{i}\sigma^{x}_{i+1} +
\frac{\sqrt{3}+2}{4}\,\sigma^{y}_{i}\sigma^{y}_{i+1} -
\sigma^{z}_{i}\Big),
\\
H_{k=1,{\rm c2}} &= \sum_{i}
\big(\sigma^{x}_{i}\sigma^{x}_{i+1} -
\sigma^{y}_{i}\sigma^{y}_{i+1} +
\sigma^{z}_{i}\sigma^{z}_{i+1}\big),
\end{align}
and for the case that the Hamiltonian takes the form of a sum over projectors:
\begin{align}
	H_{k=1,{\rm proj}}=\sum_{i}\sigma_i^x\sigma_{i+1}^x.
\end{align}
In the case of $H_{k=1,{\rm c2}}$ and $H_{k=1,{\rm proj}}$, we
discarded the unimportant shift, and rescaled the energy with a positive factor.
We note that the term $\sigma^{z}_{i}\sigma^{z}_{i+1}$ is not present in the
Hamiltonians $H_{k=1,{\rm c1}}$ and $H_{k=1,{\rm proj}}$. Therefore, they
can be solved analytically by means of a Jordan--Wigner transformation.
The Hamiltonian $H_{k=1,{\rm c2}}$ can be solved by using the Bethe ansatz.


\begin{thebibliography}{99}

\bibitem{leinaas}
J.M.~Leinaas, J.~Myrheim,
Il Nuovo Cimento B {\bf 37}, 1 (1977).

\bibitem{tsui}
D.C.~Tsui, H.L.~Stormer, A.C.~Gossard,
Phys. Rev. Lett. {\bf 48}, 1559 (1982).

\bibitem{laughlin83}
R.B.~Laughlin,
Phys. Rev. Lett. {\bf 50}, 1395 (1983).

\bibitem{MooreRead}
G. Moore and N. Read, Nucl. Phys. B {\bf 360}, 362 (1991).

\bibitem{morf}
R.H.~Morf,
Phys. Rev. Lett. {\bf 80}, 1505 (1998).

\bibitem{kitaev-chain}
A.Y.~Kitaev,
Physics-Uspekhi {\bf 44}, 131 (2001).

\bibitem{oreg}
Y.~Oreg, G.~Refael, F.~von Oppen,
Phys. Rev. Lett. {\bf 105}, 177002 (2010).

\bibitem{lutchyn}
R.M.~Lutchyn, J.D.~Sau, S.~Das Sarma,
Phys. Rev. Lett. {\bf 105}, 077001 (2010).

\bibitem{mourik}
V.~Mourik, K.~Zuo, S.M.~Frolov, S.R.~Plissard, E.P.A.M.~Bakkers, L.P.~Kouwenhoven,
Science {\bf 336}, 1003 (2012).

\bibitem{deng}
M. T. Deng, C. L. Yu, G. Y. Huang, M. Larsson, P. Caroff,
and H. Q. Xu, Nano Lett. 12, 6414 (2012).

\bibitem{das}
A.~Das, Y.~Ronen, Y.~Most, Y.~Oreg, M.~Heiblum, H.~Shtrikman,
Nat. Phys. {\bf 8}, 887 (2012).

\bibitem{feiguin}
A.~Feiguin, S.~Trebst, A.W.W.~Ludwig, M.~Troyer, A.~Kitaev, Z.~Wang,
M.H.~Freedman,
Phys. Rev. Lett. {\bf 98}, 160409 (2007).

\bibitem{fibintro}
S.~Trebts, M.~Troyer, Z.~Wang, A.W.W.~Ludwig,
Prog. Theor. Phys. Supp. {\bf 176}, 384 (2008).

\bibitem{lesanovsky12}
I.~Lesanovsky, H.~Katsura,
Phys. Rev. A {\bf 86}, 041601(R), (2012).

\bibitem{anyon-long-range}
S.~Trebst, E.~Ardonne, A.~Feiguin, D.A.~Huse, A.W.W.~Ludwig, M.~Troyer,
Phys. Rev. Lett. 101, 050401 (2008).

\bibitem{gils}
C.~Gils, S.~Trebst, A.~Kitaev, A.W.W.~Ludwig, M.~Troyer, Z.~Wang,
Nature Physics 5, 834 (2009).

\bibitem{schultz}
M.D.~Schulz, S.~Dusuel, J.~Vidal,
Phys. Rev. B {\bf 91}, 155110 (2015).

\bibitem{soni}
M.~Soni, M.~Troyer, D.~Poilblanc
Phys. Rev. B {\bf 93}, 035124 (2016).

\bibitem{spin1}
C.~Gils, E.~Ardonne, S.~Trebst, A.W.W.~Ludwig, M.~Troyer, Z.~Wang,
Phys. Rev. Lett. 103, 070401 (2009).

\bibitem{gils2}
C.~Gils,
J. Stat. Mech. P07019 (2009).

\bibitem{dancer}
K.A.~Dancer, P.E.~Finch, P.S.~Isaac, J.~Links,
Nucl. Phys. B  {\bf 812}, 456 (2009).

\bibitem{finch}
P.E.~Finch,
J. Stat. Mech. P04012 (2011).

\bibitem{dilute}
D.~Poilblanc, M.~Troyer, E.~Ardonne, P.~Bonderson,
Phys. Rev. Lett. {\bf 108}, 207201 (2012).


\bibitem{rr99}
N.~Read, E.~Rezayi,
Phys. Rev. B {\bf 59}, 8084 (1999).

\bibitem{maclane}
Suanders Mac Lane, \emph{Categories for the Working Mathematician}, 2nd Edition, Springer, 1978.

\bibitem{Kitaev06}
A. Y. Kitaev, Ann. Phys. {\bf 321}, 2 (2006).

\bibitem{bonderson-thesis}
P.H.~Bonderson, Ph.D.~thesis (2007) (\url{http://thesis.library.caltech.edu/2447/}).

\bibitem{Wang} Z.Wang, Topological Quantum Computation, Regional Conference
Series in Mathematics, Vol. 112 (American Mathematical Society,
Providence, RI, 2010).

\bibitem{kr88}
A.N.~Kirillov and N.Y.~Reshetikhin,
{\it Representations of the algebra $U_q(sl(2))$, $q$-orthogonal polynomials and invariants of links},
in V.G.~Kac, ed., {\it Infinite dimensional Lie algebras and groups, Proceedings of the conference held at CIRM, Luminy, Marseille}, p. 285, World Scientific, Singapore (1988).

\bibitem{pasquier}
V.~Pasquier,
Comm. Math. Phys. {\bf 118}, 355 (1988).

\bibitem{mga}
C.K.~Majumdar, D.K.~ Ghosh,
J. Math. Phys. {\bf 10}, 1388 (1969).

\bibitem{mgb}
C.K.~Majumdar, D.K.~ Ghosh,
J. Math. Phys. {\bf 10}, 1399 (1969).

\bibitem{AKLT1}
I. Affleck, T. Kennedy, E. H. Lieb, and H. Tasaki, Phys. Rev. Lett. 59, 799 (1987).

\bibitem{AKLT2}
I. Affleck, T. Kennedy, E. H. Lieb, and H. Tasaki,
Commun. Math. Phys. {\bf 115}, 477 (1988).

\bibitem{book:baxter}
R.J.~Baxter,
\textit{Exactly solved models in statistical mechanics},
Academic Press, London (1982).

\bibitem{abf}
G.E.~Andrews, R.J.~Baxter, P.J.~Forrester,
J. Stat. Phys. {\bf 35}, 193 (1984).

\bibitem{warnaar}
S.O.~Warnaar, B.~Nienhuis, K.A.~Seaton,
Phys. Rev. Lett. {\bf 69}, 710 (1992).

\bibitem{on1}
M.T.~Bachelor, B.~Nienhuis, S.O.~Warnaar,
Phys. Rev. Lett. {\bf 62}, 2425 (1989).

\bibitem{on2}
S.O.~Warnaar, M.T.~Bachelor, B.~Nienhuis,
J. Phys. A {\bf 25}, 3007 (1992).

\bibitem{zhou97}
Y.K.~Zhou, M.T.~Batchelor,
Nucl. Phys. B {\bf 485}, 646 (1995).

\bibitem{bpz84}
A.A.~Belavin, A.M.~Polyakov, and A.B.~Zamolodchikov,
Nucl. Phys. B {\bf 241}, 333 (1984).

\bibitem{byb}
P.~Di~Francesco, P.~Mathieu, and D.~S\'en\'echal,
{\it Conformal field theory},
Springer, New York (1999).

\bibitem{seaton02}
K.A.~Seaton,
J. Phys. A {\bf 35}, 1597 (2002).

\bibitem{eggert}
S.~Eggert, Phys. Rev. B {\bf 54}, R9612 (1996).

\bibitem{abb88}
F.C.~Alcaraz, M.N.~Barber, M.T.~Batchlor,
Ann. Phys. {\bf 182}, 280 (1988).

\bibitem{dvv88}
R.~Dijkgraaf, E.~Verlinde, H.~Verlinde,
Comm. Math. Phys. {\bf 115}, 649 (1988).

\bibitem{date86}
E.~Date, M.~Jimbo, T.~Miwa, M.~Okado,
Lett. Math. Phys. {\bf 12}, 209 (1986).

\bibitem{pasquier87}
V.~Pasquier,
Nucl. Phys. B {\bf 285}, 162 (1987).

\bibitem{kakashvili}
P.~Kakashvili, E.~Ardonne,
Phys. Rev. B 85, 115116 (2012).


\bibitem{peschel81}
I.~Peschel, V.J.~Emery,
Z.~Phys. B {\bf 43}, 241 (1981).

\bibitem{kurmann82}
J.~Kurmann, H.~Thomas, G.~M\"{u}ller,
Physica A {\bf 112}, 235 (1982).

\end{thebibliography}

\end{document}